\documentclass{article}

\usepackage[a4paper, margin=1.in]{geometry}
\usepackage{graphicx} 
\usepackage{amsmath}
\usepackage{tcolorbox}
\usepackage{multicol}
\usepackage{lineno}
\usepackage{url}
\usepackage[font=it]{caption}
\newtcolorbox[auto counter]{mybox}[2][]{float,title={Box~\thetcbcounter: #2},#1}
\usepackage{authblk}
\title{On-farm management strategies for reducing H5N1 transmission in dairy cattle}

\author[1,2*]{Oliver Eales}
\author[3]{Scott Ison} 
\author[3]{Rachael Gibney} 
\author[4]{Zoe Vogels} 
\author[1,2]{James M. McCaw}
\author[1,5]{Freya M. Shearer}

\affil[1]{Infectious Disease Dynamics Unit, Centre for Epidemiology and Biostatistics, Melbourne School of Population and Global Health, The University of Melbourne, Australia}

\affil[2]{School of Mathematics and Statistics, The University of Melbourne, Australia}

\affil[3]{Agriculture Victoria, Australia}

\affil[4]{Dairy Australia, Australia}

\affil[5]{Infectious Disease Ecology and Modelling, The Kids Research Institute, Perth, Australia}

\affil[*]{Corresponding author: Oliver Eales, oliver.eales@unimelb.edu.au}

\begin{document}
\maketitle

\noindent

\begin{abstract}
\noindent 
Introductions of H5N1 clade 2.3.4.4b into dairy cattle have resulted in outbreaks on dairy farms across the United States since early-2024. Outbreaks have significant consequences for animal health, result in economic losses for the dairy industry, and pose a threat to human health. Though the relative contributions of different on-farm transmission pathways remain a key uncertainty, a major route is considered to be through repeated contamination of milking stalls (i.e. the equipment and area where an individual cow is milked) due to the milking of infected animals. Here we develop mathematical models of H5N1 transmission dynamics on dairy farms, considering multiple possible transmission pathways, and identify factors that contribute to outbreak risk and on-farm interventions for mitigating risk. In particular, we demonstrate that dividing cattle into ‘milking cohorts’, with cohorts kept in separate pens or paddocks and milked in the same order every day, would be highly effective at mitigating outbreaks irrespective of the dominant transmission pathway. Cohorting cattle is most effective when implemented pre-emptively (i.e. before an outbreak) and when newly introduced cattle are kept in the final milking cohort. Additionally, we demonstrate that frequent bulk milk sample testing (e.g. weekly) would enable the rapid detection of outbreaks and implementation of reactive interventions (or scaling up of existing interventions). Our findings can support the development of management guidelines for effectively responding to H5N1 outbreaks in dairy cattle.

\end{abstract}
\pagebreak

\section{Introduction}\label{section:intro}
H5N1 influenza viruses belonging to clade 2.3.4.4b have circulated widely in bird and mammal populations since 2020 \cite{Peacock2024-bf}. An introduction of H5N1 clade 2.3.4.4b into dairy cattle from wild birds (genotype B3.13) in December 2023 resulted in outbreaks on dairy farms across the United States (US) \cite{Bellido-Martin2025-hf,Nguyen2025-bq,Caserta2024-mq}. These have had serious consequences for animal health \cite{Caserta2024-mq,Facciuolo2025-wy,Halwe2024-in}, resulting in significant economic losses for the dairy industry \cite{Morel2025-is,Pena-Mosca2025-lk}, and posed a threat to human health through spillover infections and the contamination of raw milk products. Furthermore, continued circulation of H5N1 in cattle —-- with frequent spillover into humans —-- increases the risk of an influenza virus emerging that is capable of efficient human-to-human transmission. Although human-to-human transmission has not been detected, experimental challenge studies in ferrets (the animal model predominantly used for assessing transmissibility of influenza viruses between humans) have demonstrated the virus' potential for ongoing respiratory transmission \cite{Gu2024-wk,Brock2025-yq}.\\
\indent The initial dairy farm outbreaks in early-2024 were the result of an introduction into dairy cattle of the genotype B3.13, a relatively rare genotype in wild birds at the time \cite{Peacock2024-bf}. In early 2025, two introductions of genotype D1.1 (a genotype relatively prevalent in North American wild birds by early-2025) \cite{Mostafa2025-pe, Crespo-Bellido2026-at} were detected \cite{UnknownUnknown-iv,UnknownUnknown-jm}. The potential for other genotypes to cause infection in dairy cattle have also been demonstrated by experimental inoculation \cite{Halwe2024-in}. This suggests that even if outbreaks are controlled, re-introductions may occur into the future, triggering new outbreaks in the US and potentially internationally. It is crucial that management strategies for effectively preventing and responding to H5N1 outbreaks on dairy farms are developed and adopted.\\
\indent National management guidelines in the US have broadly focused on testing to prevent the interstate transfer of infected cattle \cite{United-Stated-Department-of-Agriculture:-Animal-and-Plant-Health-Inspection-ServiceUnknown-zb}. However, modelling suggests that interstate export controls have had minimal impact on the propagation of outbreaks across the US, and that `farm-focused' interventions are required to prevent and control H5N1 outbreaks \cite{Rawson2025-hq}. In 2024 the US Department of Agriculture (USDA) commenced a nationwide voluntary Dairy Herd Status Program \cite{UnknownUnknown-vi}, in which farms can enrol for weekly testing of bulk tank milk samples for H5N1; a herd with three consecutive negative tests receives `Monitored Unaffected' herd status allowing the farm to move animals interstate without additional testing. Additionally, such frequent testing would likely allow outbreaks to be detected rapidly in enrolled farms. On 8 May 2026, 130 herds across 21 states were enrolled in the program \cite{UnknownUnknown-vi}. With 23,609 licensed dairy farms in the US in 2025 \cite{noauthor_2026-gj}, this limited enrolment may reflect privacy concerns or a lack of intervention measures available for farms to effectively reduce the risk and size of outbreaks.\\ 
\indent Here we use mathematical analysis and a stochastic simulation model of H5N1 transmission dynamics on dairy farms to identify opportunities for controlling and mitigating outbreaks. We focus on on-farm intervention and testing strategies that are practical to implement as evidenced by existing policy and guidelines. In particular, we investigate: enhanced cleaning of milking stalls (i.e. the area in which an individual cattle is milked); and the division of cattle into separate `milking cohorts', with cohorts kept in separate pens or paddocks and milked in the same order every day. We evaluate the effectiveness of measures implemented pre-emptively (i.e. before an outbreak) and the relative effectiveness of introducing the same measures reactively (following a positive test) under different bulk milk sample testing frequencies. Our findings will be useful to inform the development of management strategies for reducing H5N1 transmission in dairy cattle, and for responding to future outbreaks. 

\section{A mathematical description of outbreak propagation between farms}\label{section:prop}
\begin{figure*}[b!]
    \centering
    \includegraphics[width=0.5\linewidth]{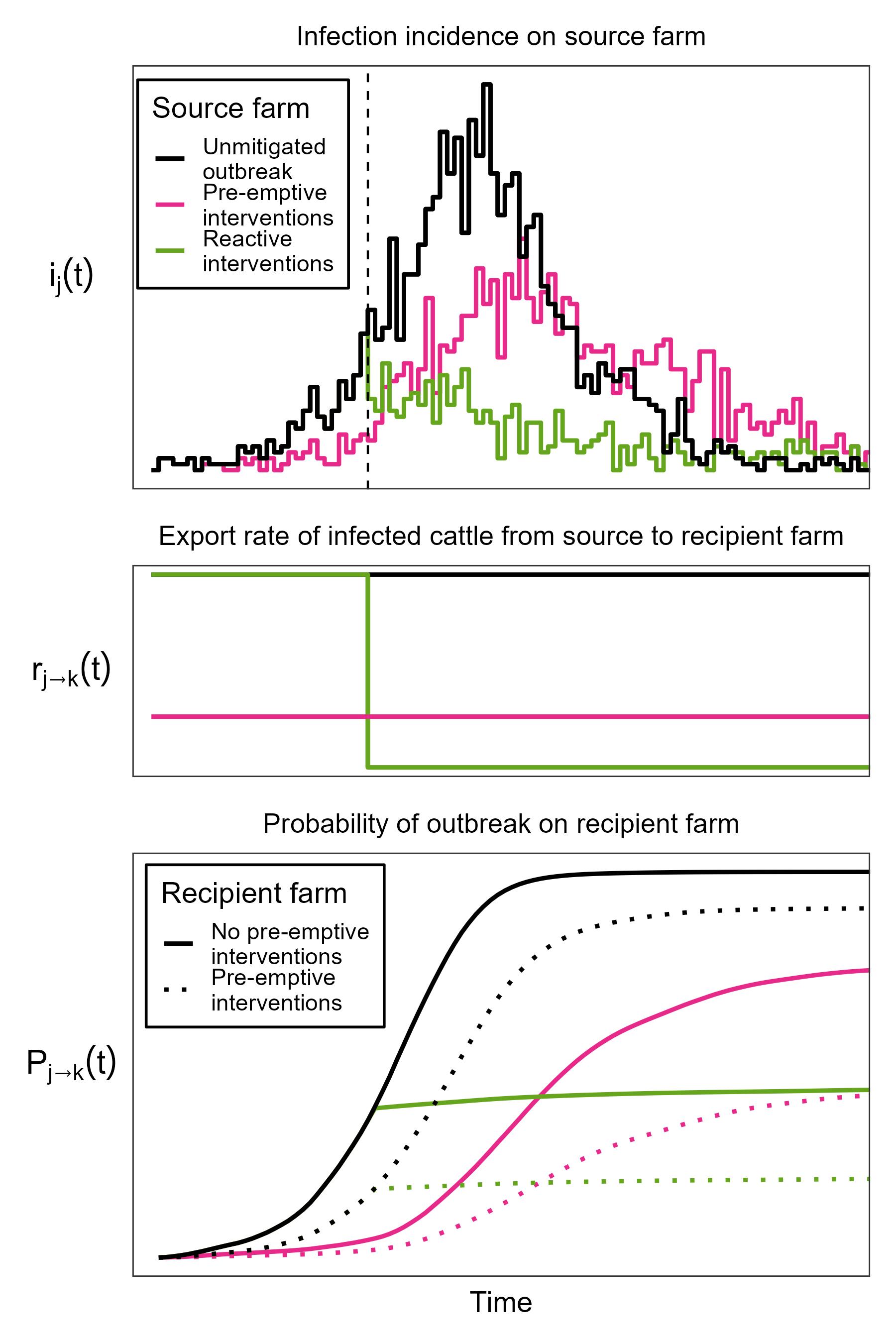}
    \caption{Graphical representation of the factors controlling the onward propagation of outbreaks. The probability of an outbreak on a source farm resulting in an outbreak on a recipient farm (bottom panel) depends on: the infection incidence of the outbreak on the source farm (top panel); the rate at which infected animals are exported from the source to the recipient farm (middle panel); and the probability that an introduced infected animal results in an outbreak (Figure \ref{fig:propagation-of-outbreaks-phi-sup}). These will be impacted by the intervention measures in place. Pre-emptive intervention measures on the source farm (pink lines) may reduce the outbreak size (infection incidence) on the source farm (top panel) or reduce the rate at which infected animals are exported (middle panel). Reactive intervention measures (green) can only be implemented following the detection of an outbreak (dashed vertical line). They may similarly reduce outbreak size on the source farm (top panel) and reduce the rate at which infected animals are exported (middle panel). Reactive intervention measures will only be implemented on farms where an outbreak has been detected, and so may be stricter than pre-emptive measures. Pre-emptive measures on the recipient farm (dotted lines), may reduce the probability an outbreak occurs following the importation of an infected animal (Figure \ref{fig:propagation-of-outbreaks-phi-sup}). In general, a combination of different pre-emptive interventions, reactive intervention, and testing strategies, may be implemented on any given farm.} \label{fig:propagation-of-outbreaks}
\end{figure*}
We begin by developing a mathematical description of outbreak propagation between farms in order to identify the factors impacting spread that could be targeted by on-farm interventions. In the US, dairy cattle are frequently transported between farms intra- and inter-state \cite{Sellman2022-as}. The movement of infected cattle --- many of which may be asymptomatic --- between dairy farms is a key driver of the propagation of H5N1 dairy farm outbreaks \cite{Nguyen2025-bq,Caserta2024-mq}. The propagation of outbreaks can be considered in terms of the probability that an outbreak on one farm (the source farm, farm $j$) directly causes an outbreak on another farm (the recipient farm, farm $k$) \cite{Yan2018-qs}. This will depend on: (1) the size of the outbreak on the source farm, characterised by the infection incidence over time, $i_j(t)$; (2) the probability that an infected animal on the source farm is transported to the recipient farm, characterised by the rate over time at which infected animals are exported from source to recipient farm, $r_{j\to k}(t)$; and (3) the probability that an infected animal introduced to the recipient farm causes an outbreak, $\phi_k(\theta)$, which will depend on the time since the animal was infected, $\theta$, as well as factors on the recipient farm (hence the subscript $k$). The probability of an outbreak on the source farm causing an outbreak on the recipient farm, $P_{j \to k}$, is then given by: 
\begin{equation}\label{eqn: probability j to k}
    P_{j \to k} = \lim_{t\to\infty} 1- \exp(-\mu_{j \to k}(t)),
\end{equation}
where 
\begin{equation}\label{eqn: mu j to k}
    \mu_{j\to k}(t) = \int_0^t r_{j \to k}(\tau)\left( \int_0^\tau \phi_{k}(\tau - \chi) i_j(\chi) \text{d}\chi \right)  \text{d}\tau.
\end{equation}
The expression being integrated over (from $0$ to $t$) in equation \ref{eqn: mu j to k} is the rate at which farm $j$ introduces an outbreak to farm $k$ over time (see Supplementary Methods). The smaller the value of $\mu_{j\to k}(t)$ the smaller the probability of a secondary outbreak at farm $k$. The expected number of secondary outbreaks arising from source farm $j$ will scale (approximately) with the sum of outbreak probabilities over all possible recipient farms, $\sum_k P_{j \to k}$ (this is only approximate as the probabilities $P_{j\to k}$ will likely not be independent). These equations highlight which factors increase or decrease the risk of a farm seeding secondary outbreaks --- the larger the farm (larger outbreaks are expected on farms with more cattle), the higher the rate at which cattle are exported from a farm, and the greater the number of recipient farms (per export farm), the greater the risk of outbreak propagation.\\
\indent Equations \ref{eqn: probability j to k}--\ref{eqn: mu j to k} also indicate where opportunities exist for preventing outbreaks or slowing the onward propagation of outbreaks. Interventions that decrease the terms in equation \ref{eqn: mu j to k} will reduce the expected number of secondary outbreaks. Existing interventions predominantly target $r_{j\to k}(t)$, the rate at which infected animals are exported. For example, testing of exported cattle can reduce $r_{j\to k}(t)$ before an outbreak has been detected, while export restrictions for farms with a detected outbreak can reduce $r_{j\to k}(t)$ (Figure \ref{fig:propagation-of-outbreaks}). Interventions could also be developed to reduce the size of outbreaks at the source farm ($i_j(t)$) and probability of outbreaks at the recipient farm ($\phi_k(\tau)$). Pre-emptive measures that reduce on-farm transmission would: (1) reduce the probability of an outbreak occurring (following the introduction of an infected animal); and (2) when an outbreak occurs, reduce the overall size of the outbreak (Figure \ref{fig:propagation-of-outbreaks}). Reactive measures that reduce on-farm transmission could also reduce the overall size of the outbreak, provided that testing strategies can identify outbreaks in a timely manner. In the following sections we will explore interventions and testing strategies that target the reduction of outbreak risk and size (i.e. $\phi_k(\tau)$ and $i_j(t)$ in equation \ref{eqn: mu j to k}).

\section{A mathematical description of on-farm transmission}\label{section:math}
A major route of transmission between dairy cattle is considered to be through the repeated contamination of milking stalls (i.e. the milking unit, feed trough, milking platform and air around which an individual cattle is milked) due to the milking of infected animals: far greater viral RNA loads have been detected in milk samples from infected animals compared to nasal swabs \cite{Caserta2024-mq,Halwe2024-in}; high infectious viral loads have been isolated from milk samples \cite{Caserta2024-mq}; and influenza A virus in milk samples pipetted onto rubber inflation liners and stainless steel --— key materials of individual milking units --— have been detected many hours after exposure \cite{Le-Sage2024-mc, Kaiser2025-zh}. It is also thought that a degree of transmission occurs without the intermediary contamination of milking stalls. For example, via contamination of the air with aerosols (i.e. direct respiratory transmission) or via contaminated wastewater (e.g. water used for washing down dairies) that may be reclaimed and used at multiple sites on the farm  increasing the presence of H5N1 in the environment \cite{Campbell2026-af}. However, the presence or relative importance of these transmission routes have not been described in the literature at the time of writing. \\
\indent To gain insight into the effect of multiple possible transmission routes we have developed a deterministic model of the on-farm transmission of H5N1, explicitly modelling the infection status of dairy cattle and contamination status of milking stalls. The model has also been adapted into a less analytically tractable, but more realistic stochastic model which we have used to perform simulation studies. We categorise cattle as either: susceptible to infection (S); exposed but not yet infectious (E); infectious (I); or recovered and not susceptible (R) (cattle develop immunity following H5N1 infection \cite{Facciuolo2025-wy}). We categorise milking stalls as either: uncontaminated (U); or contaminated and therefore infectious (V). Transitions between states are governed by a system of ordinary differential equations:  
\begin{align}
    \frac{\text{d}S}{\text{d}t} & = - \left(\frac{\alpha_f}{N_c} + \frac{\alpha_d}{A} \right) SI - \frac{\beta_I}{N_m}SV \label{eqn: deterministic system S}\\
    \frac{\text{d}E}{\text{d}t} & =  \left(\frac{\alpha_f}{N_c} + \frac{\alpha_d}{A} \right) SI + \frac{\beta_I}{N_m}SV - \sigma E \label{eqn: deterministic system E}\\
    \frac{\text{d}I}{\text{d}t} & = \sigma E - \gamma I \label{eqn: deterministic system I}\\
    \frac{\text{d}R}{\text{d}t} & = \gamma I \label{eqn: deterministic system R}\\
    \frac{\text{d}U}{\text{d}t} & = \eta V - \frac{\beta_V}{N_m}UI \label{eqn: deterministic system U}\\
    \frac{\text{d}V}{\text{d}t} & = \frac{\beta_V}{N_m}UI - \eta V \label{eqn: deterministic system V}.
\end{align}
The definitions of all parameters are provided in Table \ref{tab:main-params}. The number of cattle ($N_c$) and milking stalls ($N_m$) are also assumed to remain constant:
\begin{align}
    N_c & = S(t) + E(t) + I(t) + R(t)\label{eqn: N_c}\\ 
    N_m & = U(t) + V(t)\label{eqn: N_m}.
\end{align}
This system explicitly models two modes of transmission, direct cow-to-cow transmission and transmission via milking stalls (cow-to-milking stall-to-cow transmission). Direct cow-to-cow transmission can either be frequency dependent ($\alpha_f$ term in equation \ref{eqn: deterministic system S}) or density dependent ($\alpha_d$ term in equation \ref{eqn: deterministic system S}). Cow-to-milking stall-to-cow transmission involves two steps: (1) contamination of a milking stall by an infectious animal ($\beta_V$ term in equation \ref{eqn: deterministic system U}); and (2) infection of an animal by a contaminated milking stall ($\beta_I$ term in equation \ref{eqn: deterministic system S}). The cow-to-milking stall and milking stall-to-cow transmission rates, $\beta_V$ and $\beta_I$ respectively, can be written in terms of the milking rate ($m$) and the per milking probability of infection ($P_I$) and contamination ($P_V$):
\begin{align}
    \beta_V & = m P_V\label{eqn: beta_V}\\ 
    \beta_I & = m P_I\label{eqn: beta_I}.
\end{align}

\begin{table}[b]
    \centering
    \begin{tabular}{|c|c|}
    \hline
    Parameter& Interpretation\\
    \hline
    $\alpha_f$ & Frequency dependent cow--cow transmission rate \\
    $\alpha_d$ & Density dependent cow--cow transmission rate \\
    $A$ & Area (of cattle's paddock) \\
    $\beta_I$ & Milking stall--cow transmission rate \\
    $\beta_V$ & Cow--milking stall transmission rate \\
    $\sigma$ & Rate exposed cattle become infectious ($1/\sigma$ is the mean latent period) \\
    $\gamma$ & Recovery rate ($1/\gamma$ is the mean duration cattle remain infectiousness) \\
    $\eta$ & Decontamination rate ($1/\eta$ is the mean duration milking stalls remain contaminated)  \\ 
    $m$ & Milking rate ($1/m$ is the mean duration between a cow being milked)\\
    $P_I$ & Probability cow is infected when milked at a contaminated milking stall\\
    $P_V$ & Probability milking stall is contaminated when an infected cattle is milked at the stall\\
    \hline
    \end{tabular}
    \caption{Parameters in the mathematical description of on-farm transmission}
    \label{tab:main-params}
\end{table}

\indent The basic reproduction number, $R_0$, is the expected number of secondary infections in cattle caused by a single infected cow in an otherwise susceptible population (more formally the largest eigenvalue of the next generation matrix). For a deterministic system, when $R_0>1$ infections will increase, and when $R_0<1$ infections will die out. For the system defined by equations \ref{eqn: deterministic system S}--\ref{eqn: beta_I} $R_0$ is defined as:
\begin{equation} \label{eqn: R0}
    R_0 = \frac{ R_{0,\alpha} + \left( R_{0,\alpha}^2 + 4R_{0,\beta}^2 \right)^\frac{1}{2}}{2},
\end{equation}
where we have defined $R_{0,\alpha}$ and $R_{0,\beta}$ as:
\begin{equation} \label{eqn: R0 alpha and beta}
    R_{0,\alpha} = \frac{\alpha_f + \alpha_d\frac{N_c}{A}}{\gamma},\  \ \ \ \ \ \ \ \ 
    R_{0,\beta} = \left(\frac{m^2 P_I P_V}{\gamma \eta} \frac{N_c}{N_m} \right)^\frac{1}{2}.
\end{equation}
$R_{0,\alpha}$ is the theoretical definition of $R_0$ if no transmission is via milking stalls, whereas, $R_{0,\beta}$ is the theoretical definition of $R_0$ if all transmission is via milking stalls. The primary route of transmission between dairy cattle is considered to be via contamination of milking stalls (particularly the milking units that directly attaches to the animal to extract milk). As such $R_{0,\alpha}/R_{0,\beta}$ is small and we can use a Taylor expansion to write:
\begin{equation}
R_0 \approx R_{0,\beta}\left(1 + \frac{1}{2}\frac{R_{0,\alpha}}{R_{0,\beta}} +\mathcal{O}\left(\frac{R_{0,\alpha}}{R_{0,\beta}}^2\right) \right).
\end{equation}
The value of $R_0$ is approximately given by $R_{0,\beta}$, with any additional transmission via a mechanism other than milking stalls leading to a net increase in $R_0$. Under the assumption that all transmission is via milking stalls, we can derive the probability that infected cattle introduced to a farm lead to a major outbreak \cite{Allen2012-hz}, $P_{out,\beta}$:
\begin{equation}
    P_{out,\beta} =1- \left(\left(\frac{\gamma}{\gamma+mP_V} \right)\left(\frac{mP_I\frac{N_c}{N_m} + \eta}{mP_I\frac{N_c}{N_m}}\right)\right)^{E(0)+I(0)} \label{eqn: probs-extinction},
\end{equation}
which is equivalent to one minus the probability that the initial infections ($I(0)+E(0)$) become extinct before causing a major outbreak \cite{Allen2012-hz}.

If no transmission occurs via milking stalls ($R_0=R_{0,\alpha}$) then the probability that infected cattle introduced to a farm leads to a major outbreak is given by the (standard) expression: $P_{out,\alpha}=1-\left(1/R_{0,\alpha}\right)^{I(0)+E(0)}$.\\

\section{Factors increasing (or decreasing) outbreak risk}\label{section:factors}
We can identify factors that increase (or decrease) the risk and size of H5N1 outbreaks on a dairy farm by looking at the form of $R_{0,\beta}$ in equation \ref{eqn: R0 alpha and beta}, and $P_{out,\beta}$ in equation \ref{eqn: probs-extinction}. Farms on which cattle are milked more frequently will have greater rates of transmission (e.g. $R_{0,\beta} \propto m$). Dairy cattle are usually milked twice daily, but the frequency varies between farms, with farms regularly milking from one to three times per day \cite{Stelwagen2013-kc}. Farms may also vary their milking frequency over time depending on the lactation period of cattle or pasture abundance \cite{Stelwagen2013-kc}; during periods of higher milking frequencies farms will be at greater risk of an outbreak. Farms with a higher number of cattle per milking stall will also be at greater risk of an outbreak, with greater rates of transmission (i.e. $R_{0,\beta} \propto (N_c/N_m)^{\frac{1}{2}}$).\\
\indent We can also identify where opportunities may exist for reducing the risk and size of H5N1 outbreaks. The longer that cattle remain infectious the greater the rate of transmission (i.e.$R_{0,\beta} \propto (1/\gamma)^{\frac{1}{2}}$). If infectious cattle could be detected and then prevented from sharing milking stalls with susceptible cattle, this would effectively reduce their mean duration of infectiousness ($1/\gamma)$. Case isolation of clinical cases is unlikely to be effective unless a high proportion of infectious cattle could be detected rapidly following infection (that is, while they remain infectious) \cite{Eales2026-op}; seroprevalence studies following an outbreak on a US dairy farm suggested that approximately 25\% of infections were clinical (and thus more readily detected) \cite{Pena-Mosca2025-lk}. While subclinical cases are likely to be less infectious, their detection and isolation would contribute to outbreak control. The longer that contaminated milking stalls remain infectious the greater the rate of transmission (i.e.$R_{0,\beta} \propto (1/\eta)^{\frac{1}{2}}$). The viral concentration of influenza on milking unit rubber liners has been found to decay with a half life of approximately one hour \cite{Le-Sage2024-mc}, although there is now some evidence that the half life may be longer (based on irradiated milk samples) \cite{Kaiser2025-zh} and the number of half lives to reach zero infectiousness is not known. Interventions aimed at reducing the mean duration that units remain contaminated ($1/\eta$) would be most effective if implemented when infectiousness is highest. Contamination happens (with some probability) when an infected animal is milked --- biosecurity interventions that decontaminate units, performed at more frequent intervals would reduce the mean duration of milking stall contamination. Additionally, biosecurity interventions that reduce the per milking probability of infection ($P_I$) or contamination ($P_V$) would further reduce transmission (i.e. $R_{0,\beta} \propto (P_I P_V)^{\frac{1}{2}}$). Biosecurity protocols include pre-milking teat cleaning, post-milking teat disinfection \cite{Dairy-Australia2020-nv}, improved ventilation, and the regular cleaning of milking equipment and the milking stall feed trough \cite{Reinemann2013-nb}. The effectiveness of these measures in preventing H5N1 transmission via milking stalls has not been empirically estimated, but is expected to depend on how frequently effectively, and consistently the measures are implemented.

\section{Stochastic simulation of on-farm outbreaks}\label{section:stochastic}
To investigate the effect of potential interventions on outbreak risk and size requires some additional features to be incorporated into the model above. We extended the mathematical description above into a stochastic simulation model that more realistically describes the milking process on dairy farms (see Methods for full details). Briefly, milking is simulated as discrete-time events (i.e. milking processes start and finish at the same time each day) during which transmission between cows and milking stalls can occur. During the milking process $N_m$ cows are randomly selected and assigned to the $N_m$ milking stalls, and time is advanced by the milking time $\tau_m$ (i.e. the time to milk a single cow). This process is iterated over until all cattle have been milked. Contamination of a milking stall occurs with fixed probability upon milking of an infected animal. The probability of an animal being infected upon milking within a contaminated milking stall decays exponentially with time since contamination. Note that one could also consider a decay with the number of uses since contamination (i.e. milking of non-infectious cattle could reduce viral concentration on contaminated surfaces). The time between all other events (e.g. cow--cow transmission, recovery from infection) are determined using a Gillespie algorithm.\\
\indent
We considered three sets of parameters (see Methods) describing three transmission regimes: cow--cow transmission only; cow--milking stall--cow transmission only; and mixed mode transmission (representing predominantly cow--milking stall--cow transmission). All parameters (except for the transmission parameters) were selected to match estimates in the literature (see Methods). The transmission parameters for all three transmission regimes were selected to result in an approximate mean attack rate of 89\% (selected to align with the size of an outbreak on a farm in the US that was estimated using serological data \cite{Pena-Mosca2025-lk}) across simulations (in which an outbreak occurs) for dairy farms with 50 milking stalls ($N_m=50$) and 1000 cattle ($N_c=1000$) that are milked twice daily. This alignment between parameter sets was used to establish the same baseline scenario for all three transmission regimes, allowing the relative effect of proposed interventions to be compared fairly across those regimes.

\begin{figure}[h]
    \centering
    \includegraphics[width=1.0\linewidth]{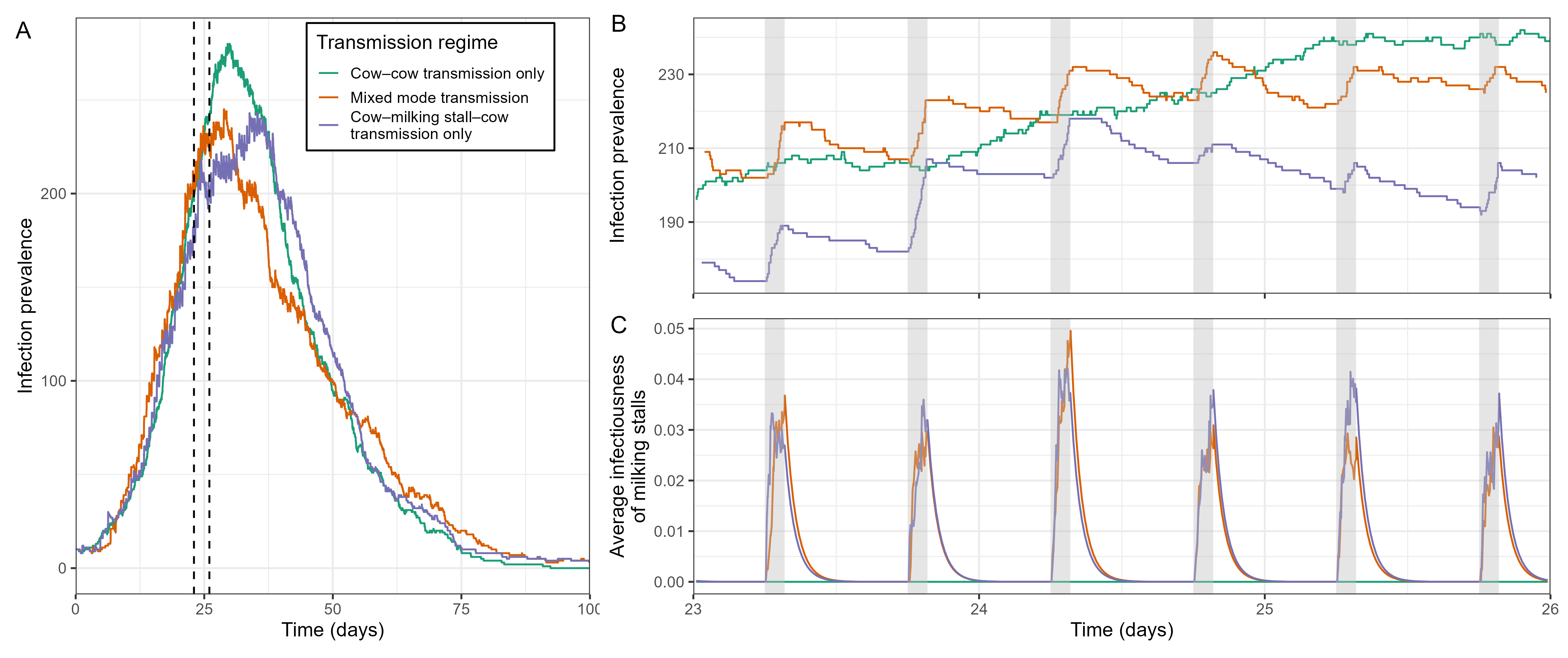}
    \caption{Example outbreak simulation for each transmission regime. (A) The infection prevalence ($N_E+N_I$) is shown over time (line) for each transmission regime (colour). (B) The time-series of infection prevalence are shown zoomed in for the period within the dashed lines in panel A. Grey regions show where the milking period occurs.
    For cow--milking stall--cow transmission only, the fact that infections can only occur during milking is clearly visible as infection prevalence only increases in these grey regions. (C) The average infectiousness of milking stalls over time (i.e. the average probability of infection if a susceptible cow is milked). The infectiousness increases during the milking period and then decays exponentially between milking periods.
    For the regime in which only cow--cow transmission occurs, the milking stalls are never infectious.}
    \label{fig:example_traj}
\end{figure}

\section{The effect of pre-emptively separating herds into milking cohorts}\label{section:pre-empt}
\begin{figure}
    \centering
    \includegraphics[width=1.0\linewidth]{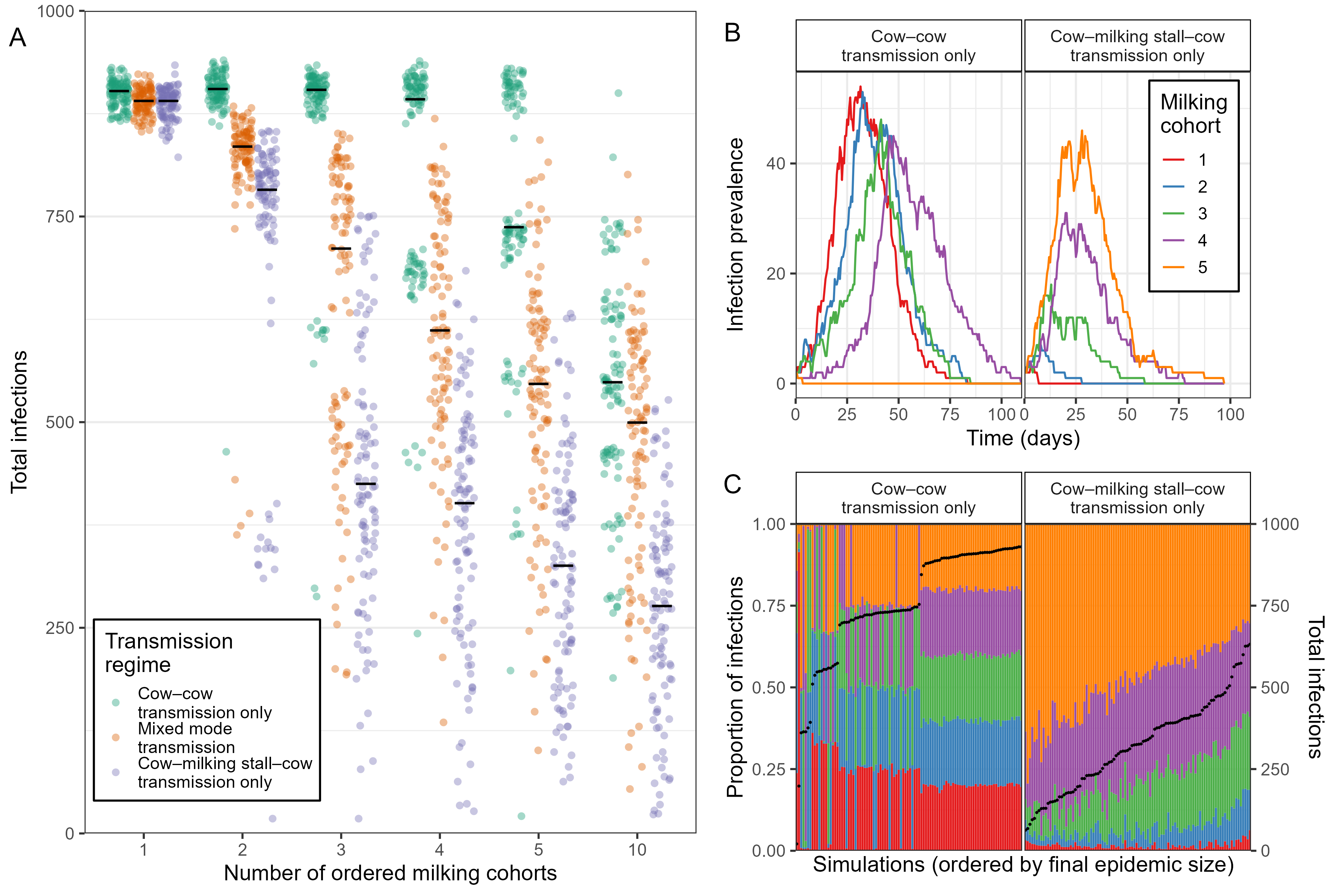}
    \caption{The effect of pre-emptively separating the dairy herd into distinct milking cohorts. (A) The total outbreak size (i.e. the total number of infections) across 100 simulations seeded with ten initial infections ($N_E(t=0)=10$, spread approximately evenly across cohorts), for three transmission regimes (colours) and for different numbers of milking cohorts (x-axis). For all scenarios the median outbreak size across 100 simulations is also shown (black line). (B) Example outbreak trajectories from simulations for five milking cohorts for the cow--cow transmission only and cow--milking stall--cow transmission only regimes. The infection prevalence ($N_E+N_I$) is shown over time (line) for each milking cohort (colour). (C) The overall distribution of infections across all simulations with five milking cohorts for the cow--cow transmission only and cow--milking stall--cow transmission only regimes. Stacked bars show the proportion of infections in each cohort (colours, left-hand y-axis) across all simulations. The simulations are ordered from smallest outbreak to largest outbreak, with the total outbreak size also plotted (points, right-hand y-axis).
    }
    \label{fig:n_cohorts}
\end{figure}

Separating cattle into distinct ordered `milking cohorts' (cohorts milked in the same order every day) prior to the introduction of infections (i.e. pre-emptively) led to a reduction in infections across all scenarios (Figure \ref{fig:n_cohorts}A). We performed 100 simulations of an outbreak --- initially seeded with 10 exposed (E compartment) cattle --- across all three transmission regimes, and for different numbers (1, 2, 3, 4, 5, and 10) of equally sized milking cohorts. The ten initial infections were distributed (approximately) uniformly across all milking cohorts. Milking cohorts were assumed to be housed in separate paddocks with no interaction between cohorts in lane ways (i.e. we assumed that no cow--cow transmission occurred between cohorts for the main analyses, even when studying the cow--cow transmission only regime), with milking cohorts milked in the same order each day. The greater the number of milking cohorts, the smaller the outbreak size (on average). The reason for this pattern differs between the three transmission regimes.

When only cow--cow transmission was assumed, outbreaks were effectively distinct between the milking cohorts (no transmission occurring between cohorts, although we relaxed this assumption in the sensitivity analysis below). As the number of milking cohorts was increased the initial number of infections seeding the outbreaks in each cohort decreased (one initial infection in each cohort for ten cohorts), which increased the probability of stochastic extinction; the reduction in overall outbreak size was thus due to a number of the cohort-level outbreaks never taking off (Figure \ref{fig:n_cohorts}).
 
When only cow--milking stall--cow transmission was assumed, the reduction in outbreak size (Figure \ref{fig:n_cohorts}A) was due to the presence of a consistent milking order of the cohorts. A hierarchy in attack rates was observed between milking cohorts; the highest attack rate was in the last milking cohort, while the smallest attack rate was in the first milking cohort (Figure \ref{fig:n_cohorts}B\&C). This was due to the rapid decay of infectiousness assumed for the contaminated milking stalls --- which was based on estimates of viral decay rates on milking unit rubber liners \cite{Le-Sage2024-mc} (we conducted a sensitivity analysis around this decay rate, see Section \ref{section:seeding}). This severely limited transmission from later cohorts to earlier cohorts as the infectiousness of contaminated milking stalls decays to near zero between milking periods. For example, cows in the last milking cohort can (with an appreciable probability) be infected by milking stalls that were contaminated by any of the preceding milking cohorts. Whereas cows in the first milking cohort are unlikely to be infected by milking stalls contaminated by other cohorts because the virus decays too quickly (i.e. between milking of the last cohort and milking of the first cohort in the subsequent milking period). 

Assuming a small degree of cow--cow transmission, with cow--milking stall--cow transmission still the dominant transmission route (i.e. the mixed mode transmission regime), also resulted in a reduction in outbreak size as the number of milking cohorts increased, but it was less dramatic than for the cow--milking stall--cow transmission only regime because of the possibility of within cohort transmission occurring outside of milking periods.

\section{The impact of seeding infections in the last cohort}\label{section:seeding}
\begin{figure}
    \centering
    \includegraphics[width=1.0\linewidth]{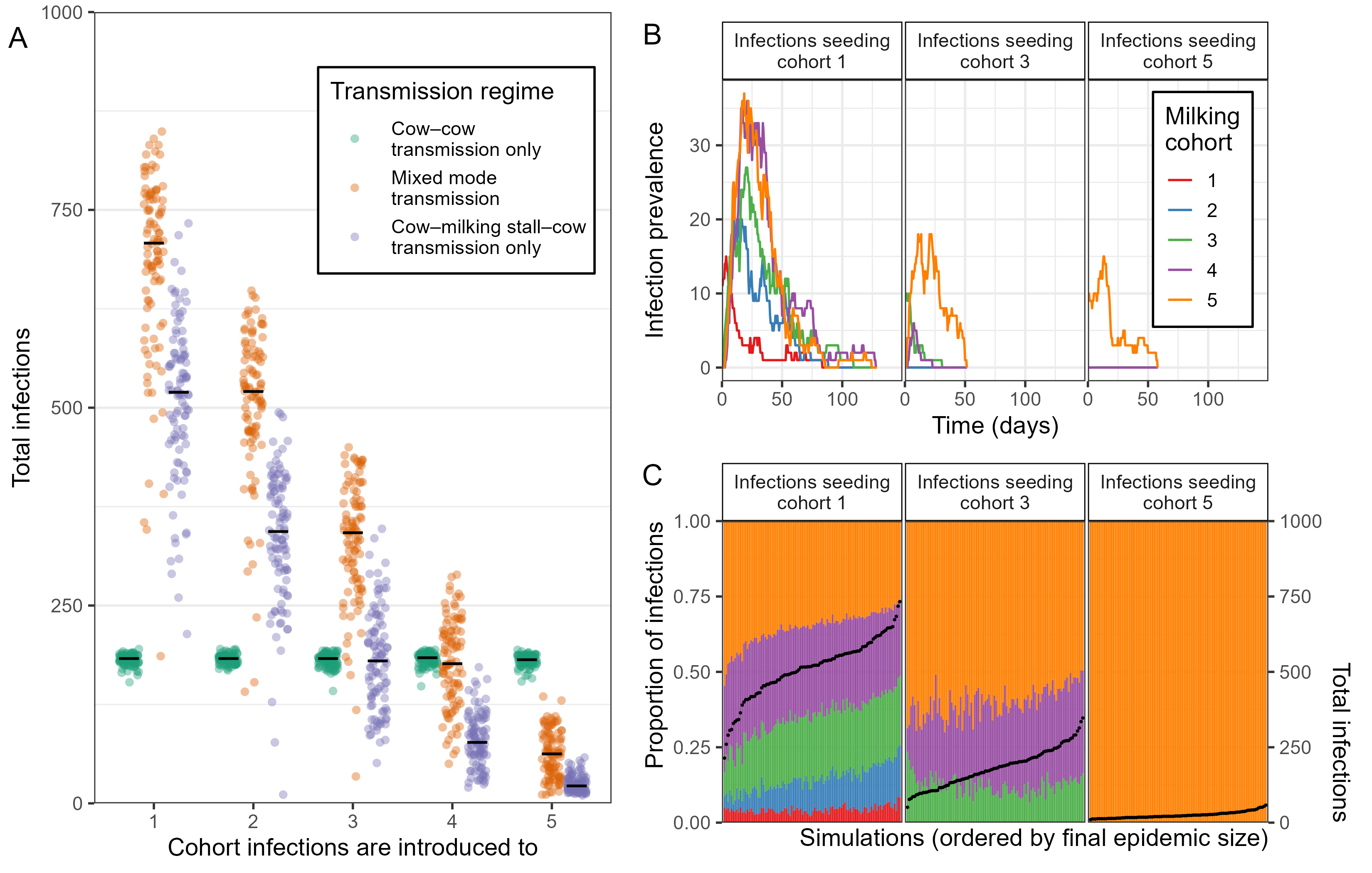}
    \caption{ The effect of the cohort in which infections are initially introduced. (A) The total outbreak size (i.e. the total number of infections) across 100 simulations seeded with ten initial infections ($N_E(t=0)=10$), for three transmission regimes (colours) and with all infections seeded into a different milking cohort (five milking cohorts in simulations). For all scenarios the median outbreak size across 100 simulations is also shown (black line) (B) Example outbreak trajectories from simulations for the cow--milking stall--cow transmission only regime, with infections seeded in cohort 1, 3, or 5. The infection prevalence ($N_E+N_I$) is shown over time (line) for each milking cohort (colour). (C) The overall distribution of infections across all simulations with infections seeded in cohort 1, 3, or 5, for the cow--milking stall--cow transmission only regime. Stacked bars show the proportion of infections in each cohort (colours, left-hand y-axis) across all simulations. The simulations are ordered from smallest outbreak to largest outbreak, with the total outbreak size also plotted (points, right-hand y-axis).
     }
    \label{fig:seed_cohorts}
\end{figure}

When investigating the effects of seeding all infections in a single cohort (i.e., non-uniform  seeding), we found that for transmission regimes including cow--milking stall--cow transmission, the later the cohort in which infections were seeded the smaller the final outbreak size. That is, the smallest outbreaks occurred when infections were seeded in the last milking cohort (Figure \ref{fig:seed_cohorts}). We performed 100 simulations of an outbreak on a dairy farm running five milking cohorts, across all three transmission regimes, and for initial infections (ten cattle in the E compartment for a single cohort) seeded in each cohort (1, 2, 3, 4, or 5). A drastic reduction in outbreak size was observed as the cohort of infection seeding was made later in the milking order. For example, under our assumptions for the cow--milking stall--cow transmission only regime, the median modelled outbreak size fell from approximately 520 cumulative infections (out of 1000 cattle, 52\% attack rate) when infections were seeded in the first cohort, to approximately 20 cumulative infections (out of 1000 cattle, 2\% attack rate) when infections were seeded in the fifth cohort. A similar effect was observed when a small degree of cow--cow transmission was assumed (mixed mode transmission regime). For both the mixed mode transmission and cow-milking stall--cow transmission only regimes, we observed the same hierarchy in outbreak size by milking cohort as when seeding infections were uniformly distributed across cohorts (see previous section). Even when all infections were seeded in the first cohort, higher attack rates were observed in the later cohorts (Figure \ref{fig:seed_cohorts}B). When infections were seeded in later cohorts, zero infections were frequently observed in earlier milking cohorts. These findings have implications for managing cattle when outbreaks are active in a region but a farm has no detected infections; newly introduced cattle should be kept in the last milking cohort to limit the possibility of infection (via milking stalls) to other milking cohorts.

For all seeding scenarios and transmission regimes that include cow--milking stall--cow transmission, the heterogeneity in attack rates between cohorts was due to the rapid decay of infectiousness assumed for contaminated milking stalls (half life of 0.87 hours), as in Section \ref{section:pre-empt}. When the decay rate was assumed to be slower (half lifes of $2\times0.87$, $3\times0.87$, and $4\times0.87$ hours, i.e. contaminated units are infectious for longer) the cohort of infection seeding had a smaller impact on the overall outbreak size (Figures \ref{fig:sens_dt1} and \ref{fig:sens_dt2}). However, these simulations assume no cleaning/active decontamination of milking stalls occurs during or between milking periods, which, if occurring, would effectively reduce the duration of milking stall infectiousness. The same benefits of employing ordered milking cohorts where the decay rate of milking stall infectiousness is fast (e.g. half life of 0.87 hours), were still achievable where the decay rate was slower (e.g. half life of $4\times0.87$ hours) under the assumption that milking stalls were cleaned and substantially decontaminated (e.g. 99\% or 99.9\% reduction in infectiousness upon cleaning) between milking periods (i.e. between milking of the last and first cohorts) (Figures \ref{fig:sens_dt2_cleaning1} and \ref{fig:sens_dt2_cleaning2}). However, if enhanced cleaning can only be implemented once per milking period and the decay rate is fast (half life of 0.87 hours), we found that enhanced cleaning was best implemented in the middle of the milking period (Figures \ref{fig:sens_cleaning1} and \ref{fig:sens_cleaning2}); this is because enhanced cleaning between the last and first milking cohort would prevent little transmission (as infectiousness of milking stalls decays to near zero over this period), for the baseline decay parameter. The optimum timing of enhanced cleaning will depend on both the effectiveness of cleaning, the decay rate of milking stall infectiousness, and the time between milking periods.  

When only cow--cow transmission was assumed, the seeding of all infections into a single cohort instead of across multiple cohorts drastically reduced the size of an outbreak. Unsurprisingly, outbreaks had approximately the same attack rate irrespective of the cohort in which seeding occurred (compare Figure \ref{fig:n_cohorts} and Figure \ref{fig:seed_cohorts}). For example, the outbreak size on a farm with five milking cohorts and uniform infection seeding resulted in an attack rate of approximately 74\% (Figure \ref{fig:n_cohorts}), whereas seeding all infections into any one of those cohorts produces an attack rate of approximately 18\% (Figure \ref{fig:seed_cohorts}). This was as expected because the outbreak now only occurred in a single milking cohort --- the cohort of infection seeding had no impact on outbreak size as transmission is assumed to be entirely independent of milking processes and cow--cow transmission occurs only within, and not between, cohorts. When a small degree of cow--cow transmission was assumed to occur between milking cohorts the overall outbreak size increased as some infections were able to seed outbreaks in other cohorts (Figure \ref{fig:sens_bct}). The higher the degree of between cohort cow--cow transmission the more cohorts experienced outbreaks and the larger the overall attack rate, highlighting the importance of biosecurity protocols that reduce the potential for transmission between cohorts. Overall, these results further support the recommendation that newly introduced cattle are kept in the last milking cohort, and that a 'sick/hospital' herd containing H5N1 affected cattle (when identified) is milked last. Although the cohort of infection seeding had no effect under the cow--cow transmission only regime, the overall outbreak size can be minimised by introducing all new cattle into the same cohort. 

\section{The impact of bulk milk sample testing frequency on outbreak detection}\label{section:BTM}

\begin{figure}
    \centering
    \includegraphics[width=1.0\linewidth]{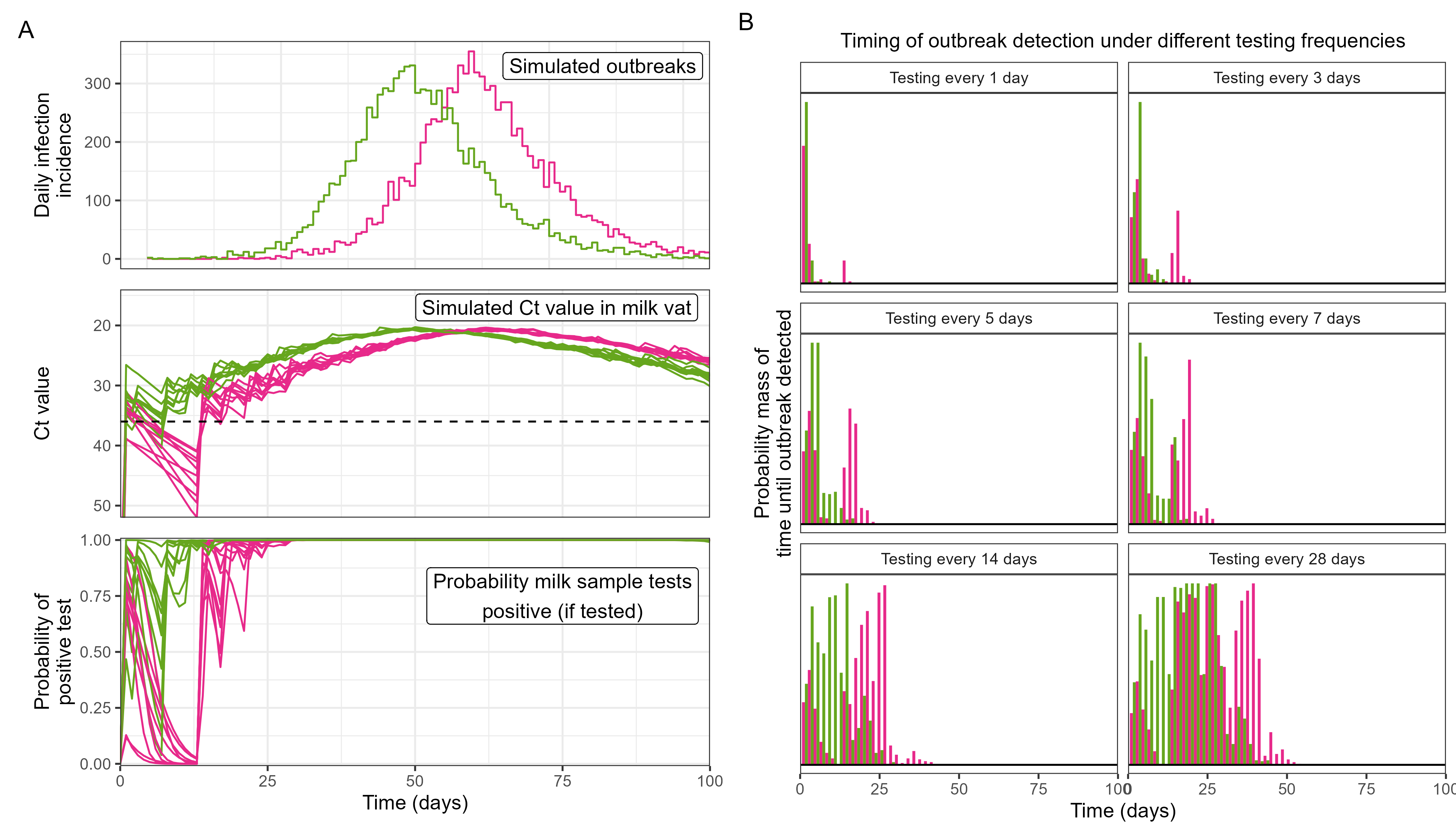}
    \caption{ Example of methods for simulating bulk milk sample testing frequencies. (A) Outbreaks are simulated and the timing of each new infection recorded. The daily infection incidence in two simulations (with $N_c=10,000$) is displayed in the top panel. For each infection simulation the expected Ct value in a bulk milk sample from the central milk vat is simulated (ten simulations per infection trajectory shown in middle panel). The probability a bulk milk sample is positive (if tested) can be calculated as the probability the measured Ct value is less than some threshold value (threshold value of 36 shown in middle panel). The probability of a positive sample for all 20 ($2\times10$) simulations is shown in the bottom panel. (B) Probability distribution of the time each simulated outbreak (colour) is detected based on the frequency bulk milk samples are tested. By performing simulations of many outbreaks the overall probability distribution across outbreak can be estimated.}
    \label{fig:testing_method}
\end{figure}

Although pre-emptively separating cattle into milking cohorts is estimated to reduce the size (and probability) of outbreaks, it may be difficult for dairy farms to implement such practices for a sustained period. Hence we also investigated the effectiveness of reactively implementing ordered milking cohorts as soon as an outbreak is detected. Since this strategy relies on early outbreak detection, we considered it in concert with different testing frequencies of bulk milk samples. We performed 100 simulations under the cow--milking stall--cow only transmission regime, for single cohort dairy farms of 100, 1000 and 10,000 cattle, with a single initial infection seeded (one cow in the E compartment). We only considered simulations in which outbreaks occurred (defined as an outbreak size greater than ten infected cattle). For each simulated outbreak we simulated the viral load and corresponding expected Ct value in a bulk milk sample (i.e. a milk sample taken from the bulk milk vat) over time (ten Ct value trajectories simulated per outbreak) using a previously published model of H5N1 viral kinetics \cite{Eales2026-op} --- conservatively assuming that infected cattle produce 25\% as much milk as non-infected cattle --- and calculated the probability of a positive test assuming a (conservative) Ct value limit of detection of 36 (see Methods) (Figure \ref{fig:testing_method}). These probabilities were then used to determine a distribution of time to outbreak detection --- assuming the outbreak is only detected by bulk milk sample testing and not due to the presence of clinically affected cattle --- across all simulated outbreaks, under different assumed testing frequencies.  

\begin{figure}
    \centering
    \includegraphics[width=1.0\linewidth]{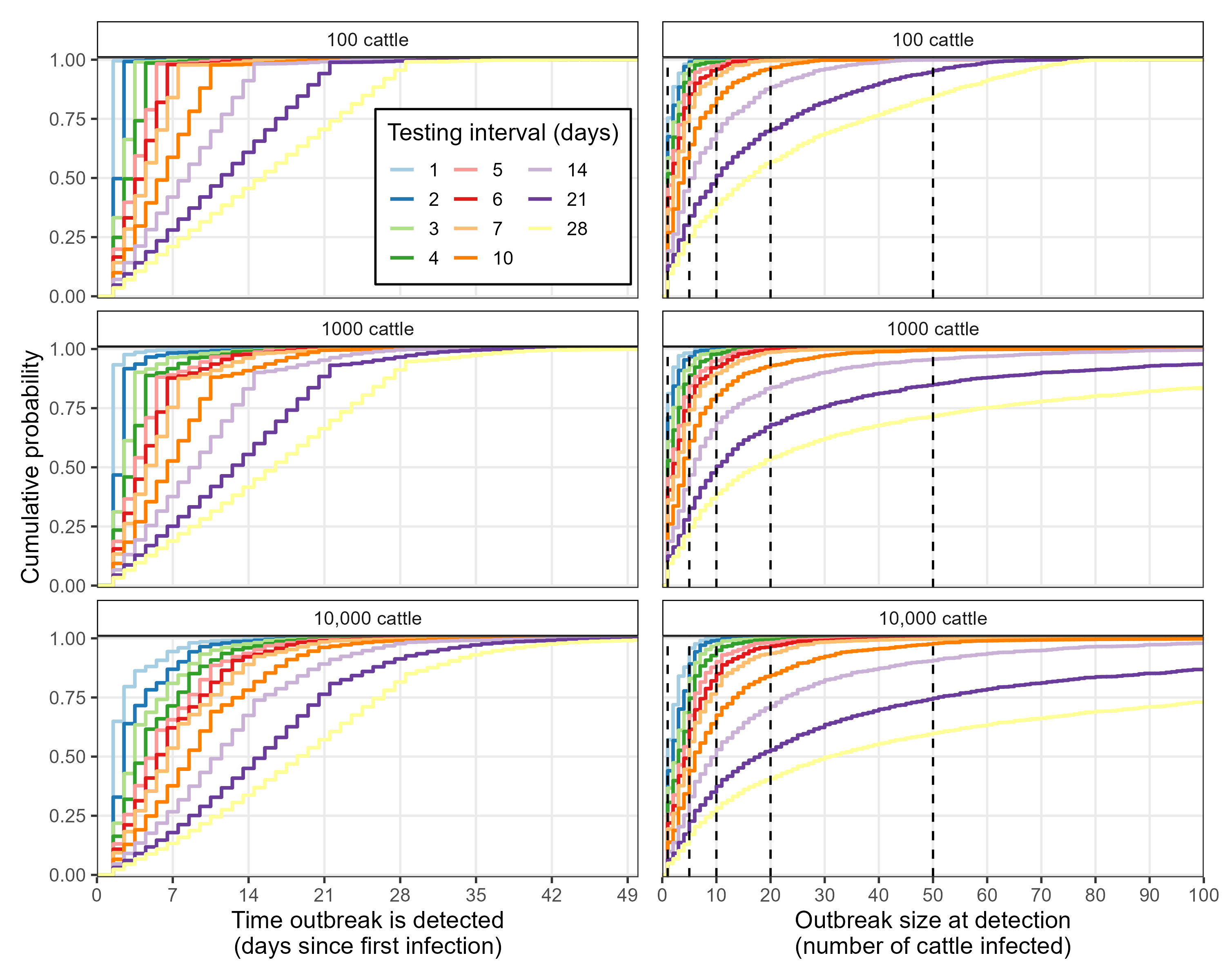}
    \caption{ The time and outbreak size at which outbreaks are detected under different bulk milk sample testing frequencies. Cumulative probability of detecting an outbreak by: (Left) time since the first infection; and (Right) total number of cattle infected (i.e. $N_E+N_I+N_R$ at time of detection). Cumulative probabilities (lines) are shown for different testing frequencies (colours). Panels show the results for simulations on different herd sizes. All simulations were performed for the cow--milking stall--cow transmission regime under for the parameters used in the main analyses. Only simulations in which an outbreak occurs (greater than ten infected cattle) were included in this analysis. Vertical dashed lines at outbreak sizes of 1, 5, 10, 20 and 50 are the thresholds at which reactive cohorting occurs in Figures \ref{fig:reactive_cohorts}, \ref{fig:reactive_cohorts_param1} and \ref{fig:reactive_cohorts_param3}. }
    \label{fig:testing_main}
\end{figure}

The time to outbreak detection was almost entirely dependent on the testing frequency, as opposed to the infection dynamics. This is because bulk milk sample testing was highly sensitive to detecting even a single infected cow in the bulk milk sample (the time taken for an infected cow's viral load to reach its maximum was approximately one day), under our conservative assumptions (see Methods). This means that on small farms (i.e. $<$100 cattle) the sensitivity of bulk milk sample testing to a single infected cow was close to 100\% (from one day post infection) and so the expected time to detect an outbreak was half the interval between tests (plus one day) (Figure \ref{fig:testing_main}). This was because the first test after introduction (which would be positive) could occur (with equal probability) after any number of days within the testing interval. 

Outbreaks were detected more slowly for larger dairy herds, but at lower infection prevalences. For some simulations on larger dairy farms (i.e. 1000 or 10,000 cattle), the outbreak was only detected on the second (or later) test --- this can be seen in the gradient changes of the cumulative probability of time to detection after the first testing interval (particularly clear for the 1,000 cattle farm in Figure \ref{fig:testing_main}). Any differences in assumptions that lead to higher sensitivity of tests or to faster outbreaks (e.g. a shorter incubation period) would shift the distribution of outbreak detection times earlier (i.e. the distribution would be closer to a uniform distribution across the testing interval). Although outbreaks were detected more slowly on larger farms, they were detected at lower infection prevalences. For example, when bulk milk sample testing was assumed to be every four weeks, 50\% of outbreaks were detected when the outbreak size was equal to or fewer than 16, 17, or 32 infected cattle (within 15, 16, and 19 days) for dairy herds of 100 (prevalence of 16\%), 1000 (1.7\%) and 10,000 (0.32\%) cattle respectively. Notably when testing was sufficiently frequent, a sizeable proportion of outbreaks were detected on a farm with only a single infected cow (Figure \ref{fig:testing_main}) --- for example, if testing was performed weekly then 37\%, 36\%, and 18\% of outbreaks were detected when there was only a single infection for dairy herds of 100, 1000, and 10,000 cattle respectively. Any reactive measures would effectively be pre-emptive under such a scenario. The relative effectiveness of reactive interventions (see Section \ref{section:react}) will directly depend on the cumulative number of infections when those interventions are implemented. As such more frequent testing would allow dairy farms to implement more effective control strategies, and reduce the disease impact on their farm. 

\section{The effect of reactively separating herds into milking cohorts upon outbreak detection}\label{section:react}
\begin{figure}
    \centering
    \includegraphics[width=1.0\linewidth]{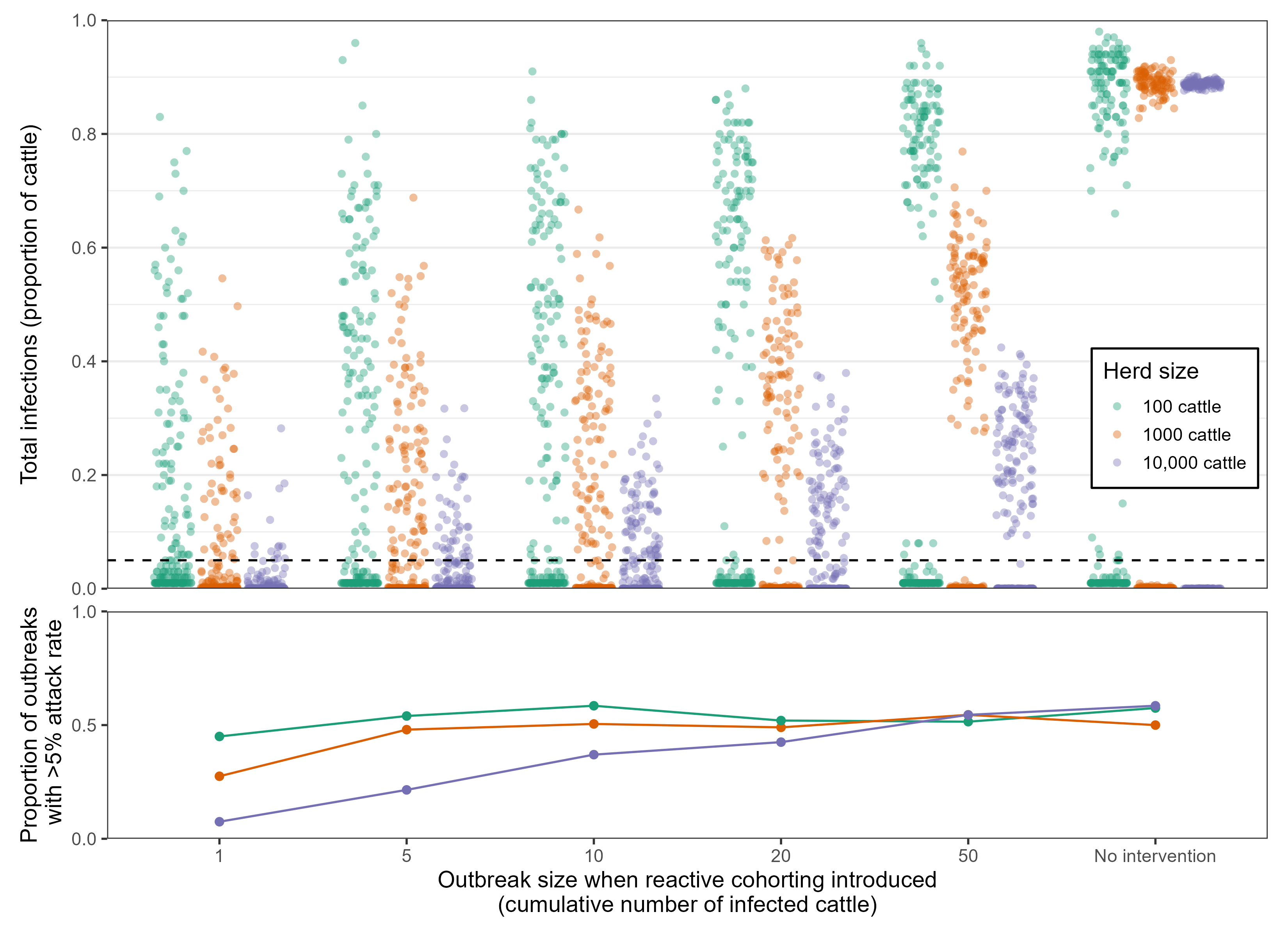}
    \caption{ The effect of reactively cohorting at different thresholds during an outbreak for the cow--milking stall--cow transmission regime. (Top) The total outbreak size (i.e. the total proportion of infected cattle) across 200 simulations seeded with one initial infection ($N_E(t=0)=1$) for the cow--milking stall--cow transmission regime (other transmission regimes shown in Figures \ref{fig:reactive_cohorts_param1} and \ref{fig:reactive_cohorts_param3}) for: different herd sizes (colours); and different thresholds of cumulative infections at which reactive cohorting (five cohorts) is performed. Note there are many simulations where outbreaks are small even when no intervention is performed; this is because the outbreak was only seeded with one initial infection and so many simulations will go extinct before an outbreak occurs, and the threshold for reactive cohorting might not be met. (Bottom). The proportion of simulations for which the outbreak size was less than 5\% (black dashed line in top panel) for the different thresholds at which reactive cohorting is performed (points) and for the different herd sizes (colours).
    }
    \label{fig:reactive_cohorts}
\end{figure}

As expected, the earlier in an outbreak that a dairy herd was reactively separated into milking cohorts (hereafter reactive cohorting), the greater the reduction in outbreak size. We performed 200 simulations of an outbreak in a dairy herd (with a single infection seeding), across all three transmission regimes, for dairy farm sizes of 100, 1000 and 10,000 cattle, assuming bulk milk sample testing detected the outbreak at different levels of cumulative infections (1, 5, 10, 20, 50, i.e. vertical dashed lines on Figure \ref{fig:testing_main}) before the dairy herd was separated randomly into five cohorts (see Methods). The largest reduction in outbreak size due to reactive cohorting was observed for the cow--milking stall--cow transmission only regime. For any threshold (cumulative infections before cohorting was implemented) the most significant reduction in attack rate was observed for the largest farms (Figure \ref{fig:reactive_cohorts}). This is not surprising because the same absolute number of infections reflects a lower infection prevalence for a larger farm. However, it is important to note that due to the high sensitivity of bulk milk sample testing it is likely that outbreaks will be detected at (relatively) similar absolute infection numbers as opposed to population prevalences. 

When a small degree of cow--cow transmission was allowed for (mixed mode transmission regime) the effectiveness of reactive cohorting was reduced (though still highly effective for the parameters assumed) (Figure \ref{fig:reactive_cohorts_param3}). When only cow--cow transmission was assumed, reactive cohorting resulted in a modest reduction in outbreak size, and attack rates of almost 20\% were frequently observed even with an outbreak detection threshold of one infected cow (Figure \ref{fig:reactive_cohorts_param1}). Overall outbreak sizes (as a proportion) for a given detection threshold were similar between farm sizes; this was because the reduction in outbreak size was due to outbreaks not occurring in some cohorts, either due to infected cattle not being allocated into those cohorts (when the reactive cohorting was implemented), or stochastic extinction occurring after infected cattle were allocated.    

\section{Discussion}\label{section:discussion}
We have developed mathematical models of H5N1 transmission dynamics on dairy farms and used them to identify factors that may contribute to outbreak risk and on-farm interventions for mitigating this risk. Previous models of H5N1 transmission have focussed on between farm transmission dynamics \cite{Rawson2025-hq}, and so have not explicitly considered the role of milking stalls in the transmission process (a valid simplifying assumption for modelling studies not investigating the effects of on-farm interventions). Here we explicitly modelled the milking process and transmission between dairy cattle through contamination of milking stalls (cow--milking stall--cow transmission), while also allowing for direct cow--cow transmission. While the relative contributions of different transmission pathways remains a key uncertainty \cite{Campbell2026-af}  --- neither transmission to cows via contaminated milking equipment (included in cow--milking stall--cow transmission in our model) nor close contact with infected animals (cow--cow transmission in our model) have been achieved under experimental conditions \cite{Lee2025-hs} --- we have identified intervention strategies that would be effective irrespective of the transmission regime. We demonstrated that separating the herd into milking cohorts, where milking cohorts are kept in separate paddocks and milked in the same order every day, reduces the risk of occurrence and size of outbreaks. Cohorting was particularly effective when outbreaks were seeded in the last cohort, suggesting that during active outbreaks newly introduced cattle should be kept in the last milking cohort.
 
Many key epidemiological parameters that determine on-farm transmission dynamics are highly uncertain and likely to vary between dairy farms and locations. Our mathematical analyses are able to highlight where interventions can be targeted to alter the ``effective" value of these parameters to reduce transmission. However, unless appropriate epidemiological data are available to robustly infer key parameters, analyses such as ours are not able to precisely estimate the expected magnitude of reductions in transmission. We have presented multiple combinations of intervention strategies that would be (theoretically) effective at reducing transmission irrespective of specific epidemiological conditions. More detailed epidemiological investigations during outbreaks could help better quantify parameters, and thus refine estimates of the effectiveness of different intervention strategies.

The duration of milking stall infectiousness is an influential but highly uncertain parameter. This parameter is informed by experimental studies that measure the viral titre of influenza virus in milk samples on surfaces over time. Variable rates of viral decay have been reported depending on surface material (e.g. stainless steel, rubber liners), temperature \cite{Le-Sage2024-mc, Kaiser2025-zh} and study. We assumed a rapid decay in milking stall infectiousness (half life of approximately 1 hour, towards the faster end of viral decays observed at $>20^\circ C$) based on the decay of influenza virus in milk samples on rubber liners (the material at the milking unit that attaches to an animal's teats) \cite{Le-Sage2024-mc}. This assumes that the decay of infectiousness of the milking stall (including the feed trough, air or other surfaces that may contribute to transmission) is similar to the decay of infectiousness of the milking unit. The model also assumes that milking stall infectiousness is proportional to the concentration of influenza virus in milk samples; a non-linear relationship is also entirely plausible (e.g. only a small amount of virus, or a threshold amount of virus, might be needed for infection to occur \cite{Lee2025-hs}). Additionally, our decay function assumes that the rate of infectiousness decays entirely due to degradation over time. It is also possible that viral concentration decays due to repeated use (i.e. milk from non-infected cattle dilutes viral concentration), or due to standard biosecurity protocols (e.g. cleaning of equipment between uses). Despite the uncertainty in the duration of milking stall infectiousness and the specific transmission regime, we have demonstrated that even when milking stalls remain infectious for longer than our decay function, the benefits of employing ordered milking cohorts will stand if effective disinfection occurs between milking periods (mimicking the effects of a faster decay rate). As one would expect, the more cleaning of milking stalls the lower the rate of transmission, but if resources (e.g. time) are constrained, limiting the frequency of cleaning, the optimum timing will depend on: the effectiveness of disinfection; the decay rate of milking stall infectiousness; and how often and when cattle are milked. In our analysis (half life of 0.87 hours, five cohorts, with milking twice daily, and cleaning once per milking period) cleaning was most effective at preventing H5N1 transmission when performed between the second and third cohorts. Although we note that cleaning would still be required following the end of each milking period to prevent bacterial growth in milk residue between milking periods.

We have developed a model to capture the salient features of on-farm transmission dynamics, but numerous simplifications had to be made for model tractability. Cow--cow transmission rates were assumed to be constant over time, but in reality will likely vary over time; for example, there may be increased opportunities for cow--cow transmission when cows are in closer contact as they move through lane ways before and after milking. We have also only considered farms with a single milking parlour, and large farms may have multiple milking parlours. In such instances ensuring that individual cattle visit the same milking parlour every milking would reduce routes of transmission, decreasing outbreak risk and size. Furthermore, observing biosecurity protocols when workers or equipment move between cattle using different milking parlours would effectively split the farm into multiple smaller farms, reducing outbreak size. Finally, we did not consider the possibility of milking orders within herds. Due to the social dynamics of cattle within a herd, a degree of natural milking order is typically observed (e.g. certain cows will always present for milking first). Even when such natural ordering occurs, further dividing a herd into ordered milking cohorts would likely increase ordering overall, and thus reduce outbreak risk and size.

We demonstrated that bulk milk sample testing is highly sensitive to detecting an outbreak of only a few infections, and so frequent testing (e.g. weekly) would allow outbreaks on farms to be rapidly detected. The nationwide (US) voluntary Dairy Herd Status Program \cite{UnknownUnknown-vi}, in which farms can enrol for weekly testing of bulk milk samples, reported limited enrolment over the course of the US outbreaks \cite{United-States-Department-of-Agriculture:-National-Agricultural-Statistics-ServiceUnknown-yo}. We have demonstrated that reactively separating cattle into milking cohorts can be highly effective at reducing the size of outbreaks, as long as outbreaks are detected rapidly. Effective reactive interventions, that become possible when undertaking bulk milk sample testing, may encourage higher enrolment rates in future monitoring programs and larger benefits for the industry. Higher enrolment rates would then also reduce between farm transmission. For example, testing measures in the United States, aimed at preventing movement of infected cattle, included required testing of a subset (up to a maximum of 30) of cattle being transported. However, the median size of dairy cattle shipments in the United States is 48 cattle \cite{Cabezas2021-ij}; for shipments larger than 48 with a single infected cow, the probability of testing that infected cow would be less than 62.5\% (30/48) whereas our results suggest that a farm conducting frequent bulk milk sample testing would be highly unlikely to export infected cattle. The exact probability of a farm exporting infected cattle would depend on the farm's size and the rate the farm exports cattle.

\section{Conclusions}\label{section:conclusion}
We have used mathematical methods to quantify the effect of different interventions for mitigating outbreaks of H5N1 influenza on dairy farms. Our findings may be considered in the development of management guidelines to inform H5N1 outbreak response strategies on dairy farms. In the event of active influenza H5N1 outbreaks affecting cattle in a region, pre-emptive interventions can be taken on infection-free farms to reduce the risk of outbreaks, and the size of outbreaks should they occur, including: pre-emptively separating herds into ordered milking cohorts; placing newly introduced cattle into the final milking cohort; regular enhanced cleaning of milking stalls; and observing strict biosecurity protocols when workers or equipment move between milking cohorts (e.g. decontamination or cleaning). Additionally, frequent bulk milk sample testing (e.g. weekly) would enable the rapid detection of outbreaks and implementation of reactive interventions (or scaling up of existing  interventions). All interventions explored in our analysis should be practical to implement, as evidenced by existing policy or practice. For example, many dairy farms routinely run multiple `mobs' (e.g. to separate cattle by A2/A1 milk type, or to quarantine mastitis-affected cattle), and milking them in a consistent order would likely require minimal additional coordination. Our findings can directly inform the development of management guidelines for effectively responding to future H5N1 outbreaks in dairy cattle.

\pagebreak

\section*{Methods}

\subsection*{Propagation of outbreaks}
The probability of an outbreak on farm $j$ seeding an outbreak on farm $k$ by time $t$ is given by \cite{Yan2018-qs}:
\begin{equation}
    P_{j \to k}(t_o<t) =  1- \exp(-\mu_{j \to k}(t)),
\end{equation}
where $t_o$ is the time an outbreak is seeded, and $\mu_{j \to k}(t)$ is the time varying rate at which farm $j$ introduces an outbreak to farm $k$ over time. This time-varying rate, can be expressed as:
\begin{equation}
    \mu_{j\to k}(t) = \int_0^t r_{j \to k}(\tau)\left( \int_0^\tau \phi_{k}(\tau - \chi) i_j(\chi) \text{d}\chi \right)  \text{d}\tau,
\end{equation}
where: $i_j(t)$ is the infection incidence on farm $j$ over time; $r_{j \to k}(t)$ is the rate at which infected animals are exported from farm $j$ to farm $k$ over time; and $\phi_k(\tau)$ is the probability that an infected animal introduced to farm $k$ causes an outbreak, as a function of the time since infection. The probability density function for the time at which an outbreak on farm $j$ seeds an outbreak on farm $k$ is then given by:
\begin{equation}
    P_{j \to k}(t_o=t) =  \exp(-\mu_{j \to k}(t)) \frac{\text{d}\mu_{j \to k}(t)}{\text{d}t}
\end{equation}
and the overall probability that farm $j$ seeds an outbreak on farm $k$ is given by:
\begin{equation}
    P_{j \to k} = \lim_{t\to\infty} 1- \exp(-\mu_{j \to k}(t))
\end{equation}
The expected number of secondary outbreaks cause by farm $j$ will scale (approximately) with the sum of outbreak probabilities,
\begin{equation}
    \sum_k P_{j \to k}.
\end{equation}
This expression is only approximate as the probabilities $P_{j \to k}$ will likely not be independent.

\subsection*{Stochastic model of H5N1 transmission}
We simulate on-farm outbreaks using a stochastic model of on-farm transmission dynamics. The model considers two modes of transmission: cow--cow transmission, that is simulated using a Gillespie algorithm; and cow--milking stall--cow transmission, that is simulated at discrete times using time-varying probabilities (see below). The model defines a farm as a number of individual cattle and milking stalls. Cattle are divided into separate milking cohorts (i.e. separate pens or paddocks, that are milked in a consistent order). Milking stalls could analogously be divided into separate milking parlours, with different cohorts milked at different milking parlours. However, we only consider one milking parlour in this analyses; if there was more than one and all cohorts only went to a single parlour consistently then the structure would be analogous (in transmission terms) to two farms.

\subsubsection*{Cow--cow transmission model}
Individual cattle are either: susceptible to infection (S); exposed but not yet infectious (E); infectious (I); or recovered and not susceptible (R). Cattle are divided into distinct milking cohorts, which are kept in separate paddocks (subscript $i$). Outside of milking events (next section), transitions between states are governed by a system of ordinary differential equations: 
\begin{align}
    \frac{\text{d}S_i}{\text{d}t} & = - \lambda_i S_i  \label{eqn: stochastic system S}\\
    \frac{\text{d}E_i}{\text{d}t} & =   \lambda_i S_i - \sigma E_i \label{eqn: stochastic system E}\\
    \frac{\text{d}I_i}{\text{d}t} & = \sigma E_i - \gamma I_i \label{eqn: stochastic system I}\\
    \frac{\text{d}R_i}{\text{d}t} & = \gamma I_i \label{eqn: stochastic system R}.
\end{align}
With transition events and times simulated using a Gillespie algorithm until the next milking event. Here, $\lambda_i$ is the force of infection on milking cohort $i$. For the main analyses the force of infection takes the form $\lambda_i=\alpha_1 \frac{I_i}{A_i}$. Where $\alpha_1$ is the density dependent transmission rate and $A_i$ is the area of the paddock (which we assume is constant between cohorts). In addition we run sensitivity analyses allowing there to be some degree of between cohort cow--cow transmission,
\begin{equation}
    \lambda_i=\alpha_1 \frac{I_i}{A_i} + \alpha_0 \frac{\sum_jI_j}{\sum_jA_j},
\end{equation}
where the summations are over all cohorts $j$. This allows there to be some additional (equal) degree of transmission between all cattle (mediated by the area of the entire farm). In all analyses we assume $\sigma = 1/1.1$ (1.1 days is the time until minimum Ct value (peak viral load) is reached in Eales et al. \cite{Eales2026-op}) and $\gamma=1/7.8$ (7.8 days is the average duration of infectiousness inferred in Eales et al. \cite{Eales2026-op}.

\subsubsection*{Cow--milking stall--cow transmission}
A milking stall becomes contaminated when an infected cow is milked at the milking stall with probability $P_{c\to m}$. Milking stall--cow transmission occurs when a susceptible cow is milked at a previously contaminated milking stall, with probability $P_{m\to c}(t)$. We assume that this probability decays exponentially over time:
\begin{equation}
    P_{m\to c}(t)=P_{m\to c}(0) \times \exp(-(t-t_{con})\zeta)
\end{equation}
with decay rate $\zeta$. 

In the main analyses we assume that $\zeta=1.15$ per hour. This was the approximate decay rate of H1N1 influenza A virus in milk on rubber liners over 5 hours (see Figure 1C and supporting data for Le Sage et al 2024 \cite{Le-Sage2024-mc}); the same study found similar decay rates between influenza A H5N1 and H1N1 over the first hour, but did not collect data over a longer period for H5N1. The decay rate for H5N1 in milk on rubber liners over one hour was approximately 1.4 per hour (see Figure 1A and supporting data for Le Sage et al 2024 \cite{Le-Sage2024-mc}). This reflects only a slightly shorter half-life (0.7 hours) than that used in the main analyses (0.87 hours). A recent study has estimated a longer half life of H5N1 in irradiated milk on rubber (approximately 3.3 hours at $22^\circ C$ and 12 hours at $4^\circ C$) \cite{Kaiser2025-zh} and so we also considered sensitivity analyses with slower decay rates for milking stall infectiousness. The probabilities, $P_{m\to c}(0)$ and $P_{c\to m}$ were calibrated, with full details given in the Simulation studies section below. In addition we run sensitivity analyses in which $P_{m\to c}(t)$ decreases by a proportion $\eta$ every time enhanced cleaning occurs (see Table \ref{tab:sims_table} for the timings of cleaning and values of $\eta$ investigated).

\subsubsection*{Timing of milking events}
Cow--milking stall--cow transmission is dependent on the timing of milking events. The timing of all milking periods (the period over which all cattle are milked) is defined at the beginning of the simulation. For all simulations we assume that cattle are milked twice daily over two milking periods starting at 0600 and 1800 (i.e. 12 hours apart). During the milking periods individual milking events (i.e. cow and milking stall pairings) are determined systematically. $N_m$ cows are randomly selected from the first cohort and paired with the $N_m$ milking stalls. Potential cow--milking stall--cow transmission events are probabilistically simulated (see above). Time is advanced by $\tau_m$ the assumed time taken for each cow to be milked (assumed to be 5 minutes throughout); during this period Gillespie events (i.e. cow--cow transmission, recovery) can occur as in the above section. The cows that have just been milked are then transferred to an empty paddock (or the same area as their original paddock), and the next $N_m$ cows are randomly selected. This is repeated until all cows in the cohort have been milked and then repeated for each of the cohorts. We assume there is no additional time between each milking cohort (i.e. as soon as first milking cohort is completed the second cohort begins). 

\subsubsection*{Simulation studies}
We perform simulation studies to explore the impact of intervention measures and parameter values. For most of the main analyses we consider three sets of transmission parameters describing three transmission regimes: cow--cow transmission only; cow--milking stall--cow transmission only; and mixed mode transmission (representing predominantly cow--milking stall--cow transmission). The transmission parameters are calibrated so that simulated outbreaks (initialised with $N_E=10$) on dairy farms with $N_c=1000$, $N_m=50$, a single milking cohort, and milking occurring twice per day (at 0600 and 1800) average to approximately 89\% \cite{Pena-Mosca2025-lk}. The transmission parameter values used in these simulations are provided in Table \ref{tab:params}. Full details for all simulations performed are provided in Table \ref{tab:sims_table}.

\begin{table}
    \centering
    \begin{tabular}{|c|c|c|c|}
    \hline
    Parameter (unit) & Cow--cow & Mixed-mode & Cow--milking stall--cow\\
    & transmission &  transmission &transmission \\
    \hline
    $\alpha_0$ ((individuals$\times$days)$^{-1}$) & 0 & 0 & 0 \\
    $\alpha_1$ ((individuals$\times$days)$^{-1}$) & 0.34 & 0.07 & 0 \\
    $P_{c\to m}$ & 0 & 0.16 & 0.18 \\
    $P_{m\to c}(0)$ & 0 & 0.16 &  0.18 \\
    \hline
    \end{tabular}
    \caption{Baseline transmission parameters for each transmission regime.}
    \label{tab:params}
\end{table}

\subsection*{Bulk milk sample testing}
We estimate the time it takes to detect an outbreak under different frequencies of bulk milk sample testing (i.e. testing a sample of milk from the central milk vat). To do this we: simulate outbreaks (see Table \ref{tab:sims_table}) on dairy farms of 100, 1000 and 10,000 cattle; simulate Ct value trajectories in bulk milk samples over time for each outbreak; calculate the probability of a positive test given the Ct value of the bulk milk sample; and finally, simulate different testing frequencies and compute when the first positive test occurs.   

\subsubsection*{Ct value model}
We simulate the H5N1 viral kinetics in milk samples of infected cattle using a previously parametrised model \cite{Eales2026-op}. The model assumes that the Ct value of a milk sample from one infected cattle is given by:
\begin{align}
    \text{Ct}_i(t_{inf}) &= P_i - r_i \times (t_{inf} - \tau_i) \quad  \text{for} \quad t_{inf}<\tau_i\\
    &= P_i + d_i \times (t_{inf} - \tau_i) \quad  \text{for} \quad t_{inf}\geq\tau_i
\end{align}
Where $P_i$ is the minimum Ct value reached, $\tau_i$ is the time since infection ($t_{inf}$) at which the minimum is reached, $r_i$ the viral growth rate (decrease in Ct) before the minimum Ct value is reached, and $d_i$ the viral decay rate (increase in Ct) after the minimum Ct value is reached. An individual cow's Ct value in milk is thus governed by their time of infection and four parameters: $P_i, \tau_i, r_i, d_i$. For each infected cow we draw parameter values from the population-level distribution using the median population-level parameter estimates from the previously published study \cite{Eales2026-op}. 

The Ct value of bulk milk samples is calculated by combining the individual-level Ct values of infected cattle. Ct value is converted to virus concentration in a milk sample ($TCID_{50}$ equivalent/ml) using the standard curve: $\text{Ct}=44.843-1.736\times \log(x)$ \cite{Facciuolo2025-wy}. The virus concentration in the bulk milk sample is then calculated by averaging over all cattle assuming conservatively that each infected cow produces 25\% of the milk produced by a non-infected cow. The virus concentration is then converted back to a Ct value using the standard curve above.

\subsubsection*{Probability of testing positive}
The probability of a bulk milk sample testing positive given its Ct value is defined as the probability that the Ct value is less than a threshold value. We chose the threshold value conservatively to be 36, though some studies have used detection sensitivities of higher Ct values (lower viral loads). We calculate the probability that the Ct value is less than the threshold value assuming that the measured Ct value will be normally distributed around its expected value with standard deviation taken from the model in Eales et al \cite{Eales2026-op}. 

\subsubsection*{Simulating testing frequencies}
We simulate different testing frequencies during the course of an outbreak and determine the time at which the first positive test occurs, given the probability of a bulk milk sample testing positive over time. We assume that testing is performed at the same time on any given day (mid-day). For any testing interval it is equally possible that the first test occurs on any day between the outbreak beginning and the first complete testing interval. For example, if testing was performed every 10 days then there is an equal probability of the first test being on any day from day 0 to day 9 of the outbreak. For each possible day of first test we simulate testing beginning on that day and being performed at regular intervals, with test results determined using the probability of a positive test.

For each infection trajectory (for outbreaks that do not immediately become extinct), we simulate 10 Ct value trajectories and calculate the probability of a positive test. We then perform 10 simulations of bulk milk sample testing for each testing frequency, and for each possible day of first test. We then consider the probability distribution over all simulations of the time to first positive test (and the number of infections prior to the first positive test).

\section*{Code availability}
All code supporting this paper is available at \url{https://github.com/Eales96/H5N1-transmission-dynamics/}

\section*{Acknowledgements}
This research is supported by the Livestock Biosecurity Fund -- Cattle Compensation Fund (CCF 25.19). OE is supported by a University of Melbourne McKenzie Fellowship and Andrew Sisson support package. FMS is supported by the National Health and Medical Research Council of Australia through the Investigator Grant Scheme (Emerging Leader Fellowship, 2021/GNT2010051). JMM is supported by the Australian Research Council through the Laureate Fellowship Scheme (FL240100126). This research is supported by the Australian Consortium of Epidemic Forecasting and Analytics (ACEFA), a National Health and Medical Research Council of Australia Centre of Research Excellence (2035303). FMS and JMM’s research is also supported by an Australian Research Council Discovery Project Grant (DP240102286).

\bibliographystyle{plos2015}
\bibliography{ref}

@ARTICLE{Peacock2024-bf,
  title     = "The global {H5N1} influenza panzootic in mammals",
  author    = "Peacock, Thomas and Moncla, Louise and Dudas, Gytis and
               VanInsberghe, David and Sukhova, Ksenia and Lloyd-Smith, James O
               and Worobey, Michael and Lowen, Anice C and Nelson, Martha I",
  journal   = "Nature",
  publisher = "Nature Publishing Group",
  pages     = "1--2",
  month     =  sep,
  year      =  2024,
  language  = "en"
}

@ARTICLE{Morel2025-is,
  title    = "The impact of {H5N1} on {US} domestic and international dairy
              markets",
  author   = "Morel, Guillaume and Pham, Anh and Morgenstern, Christian and
              Hicks, Joseph and Rawson, Thomas and Fan, Victoria and Edmunds, W
              John and Forchini, Giovanni and Hauck, Katharina",
  journal  = "Social Science Research Network",
  month    =  jan,
  year     =  2025
}

@ARTICLE{Le-Sage2024-mc,
  title     = "Persistence of influenza {H5N1} and {H1N1} viruses in
               unpasteurized milk on milking unit surfaces",
  author    = "Le Sage, Valerie and Campbell, A J and Reed, Douglas S and
               Duprex, W Paul and Lakdawala, Seema S",
  journal   = "Emerg. Infect. Dis.",
  publisher = "Centers for Disease Control and Prevention (CDC)",
  volume    =  30,
  number    =  8,
  pages     = "1721--1723",
  month     =  aug,
  year      =  2024,
  language  = "en"
}

@ARTICLE{Bellido-Martin2025-hf,
  title     = "Evolution, spread and impact of highly pathogenic {H5} avian
               influenza A viruses",
  author    = "Bellido-Martín, Beatriz and Rijnink, Willemijn F and Iervolino,
               Matteo and Kuiken, Thijs and Richard, Mathilde and Fouchier, Ron
               A M",
  journal   = "Nat. Rev. Microbiol.",
  publisher = "Springer Science and Business Media LLC",
  pages     = "1--16",
  month     =  may,
  year      =  2025,
  language  = "en"
}

@ARTICLE{Halwe2024-in,
  title     = "{H5N1} clade 2.3.4.{4b} dynamics in experimentally infected
               calves and cows",
  author    = "Halwe, Nico Joel and Cool, Konner and Breithaupt, Angele and
               Schön, Jacob and Trujillo, Jessie D and Nooruzzaman, Mohammed and
               Kwon, Taeyong and Ahrens, Ann Kathrin and Britzke, Tobias and
               McDowell, Chester D and Piesche, Ronja and Singh, Gagandeep and
               Pinho Dos Reis, Vinicius and Kafle, Sujan and Pohlmann, Anne and
               Gaudreault, Natasha N and Corleis, Björn and Ferreyra, Franco
               Matias and Carossino, Mariano and Balasuriya, Udeni B R and
               Hensley, Lisa and Morozov, Igor and Covaleda, Lina M and Diel,
               Diego and Ulrich, Lorenz and Hoffmann, Donata and Beer, Martin
               and Richt, Juergen A",
  journal   = "Nature",
  publisher = "Nature Publishing Group",
  pages     = "1--3",
  month     =  sep,
  year      =  2024,
  language  = "en"
}

@ARTICLE{Mostafa2025-pe,
  title     = "Highly pathogenic avian influenza {H5N1} in the United States:
               recent incursions and spillover to cattle",
  author    = "Mostafa, Ahmed and Nogales, Aitor and Martinez-Sobrido, Luis",
  journal   = "Npj Viruses",
  publisher = "Springer Science and Business Media LLC",
  volume    =  3,
  number    =  1,
  pages     =  54,
  month     =  jul,
  year      =  2025,
  language  = "en"
}

@ARTICLE{Gu2024-wk,
  title     = "A human isolate of bovine {H5N1} is transmissible and lethal in
               animal models",
  author    = "Gu, Chunyang and Maemura, Tadashi and Guan, Lizheng and Eisfeld,
               Amie J and Biswas, Asim and Kiso, Maki and Uraki, Ryuta and Ito,
               Mutsumi and Trifkovic, Sanja and Wang, Tong and Babujee, Lavanya
               and Presler, Jr, Robert and Dahn, Randall and Suzuki, Yasuo and
               Halfmann, Peter J and Yamayoshi, Seiya and Neumann, Gabriele and
               Kawaoka, Yoshihiro",
  journal   = "Nature",
  publisher = "Springer Science and Business Media LLC",
  volume    =  636,
  number    =  8043,
  pages     = "711--718",
  month     =  dec,
  year      =  2024,
  language  = "en"
}

@ARTICLE{Brock2025-yq,
  title    = "Avian influenza A({H5N1}) isolated from dairy farm worker,
              Michigan",
  author   = "Brock, Nicole and Pulit-Penaloza, Joanna A and Belser, Jessica A
              and Pappas, Claudia and Sun, Xiangjie and Kieran, Troy J and Zeng,
              Hui and De La Cruz, Juan A and Hatta, Yasuko and Di, Han and
              Davis, C Todd and Tumpey, Terrence M and Maines, Taronna R",
  journal  = "Emerg. Infect. Dis.",
  volume   =  31,
  number   =  6,
  pages    =  1253,
  month    =  jun,
  year     =  2025,
  language = "en"
}

@ARTICLE{Rawson2025-hq,
  title     = "A mathematical model of {H5N1} influenza transmission in {US}
               dairy cattle",
  author    = "Rawson, Thomas and Morgenstern, Christian and Knock, Edward S and
               Hicks, Joseph and Pham, Anh and Morel, Guillaume and Murillo,
               Aurelio Cabezas and Sanderson, Michael W and Forchini, Giovanni
               and FitzJohn, Richard and Hauck, Katharina and Ferguson, Neil",
  journal   = "Nat. Commun.",
  publisher = "Nature Publishing Group",
  volume    =  16,
  number    =  1,
  pages     =  4308,
  month     =  may,
  year      =  2025,
  language  = "en"
}

@ARTICLE{Caserta2024-mq,
  title     = "Spillover of highly pathogenic avian influenza {H5N1} virus to
               dairy cattle",
  author    = "Caserta, Leonardo C and Frye, Elisha A and Butt, Salman L and
               Laverack, Melissa and Nooruzzaman, Mohammed and Covaleda, Lina M
               and Thompson, Alexis C and Koscielny, Melanie Prarat and Cronk,
               Brittany and Johnson, Ashley and Kleinhenz, Katie and Edwards,
               Erin E and Gomez, Gabriel and Hitchener, Gavin and Martins,
               Mathias and Kapczynski, Darrell R and Suarez, David L and
               Alexander Morris, Ellen Ruth and Hensley, Terry and Beeby, John S
               and Lejeune, Manigandan and Swinford, Amy K and Elvinger,
               François and Dimitrov, Kiril M and Diel, Diego G",
  journal   = "Nature",
  publisher = "Springer Science and Business Media LLC",
  volume    =  634,
  number    =  8034,
  pages     = "669--676",
  month     =  oct,
  year      =  2024,
  language  = "en"
}

@TECHREPORT{UnknownUnknown-jm,
  title       = "{APHIS} Confirms {D1}.1 Genotype in Dairy Cattle in Nevada",
  institution = "U.S. Derpartment of Agriculture: Animal and Plant Health
                 Inspection Service",
  language    = "en"
}

@TECHREPORT{UnknownUnknown-iv,
  title       = "{APHIS} Identifies Third {HPAI} Spillover in Dairy Cattle",
  institution = "U.S. Derpartment of Agriculture: Animal and Plant Health
                 Inspection Service",
  language    = "en"
}

@ARTICLE{Yan2018-qs,
  title     = "The distribution of the time taken for an epidemic to spread
               between two communities",
  author    = "Yan, Ada W C and Black, Andrew J and McCaw, James M and Rebuli,
               Nicolas and Ross, Joshua V and Swan, Annalisa J and Hickson,
               Roslyn I",
  journal   = "Math. Biosci.",
  publisher = "Elsevier BV",
  volume    =  303,
  pages     = "139--147",
  month     =  sep,
  year      =  2018,
  language  = "en"
}

@MISC{UnknownUnknown-vi,
  title        = "Dairy Herd Status Program",
  booktitle    = "U.S Department of Agriculture: Animal and Plant Health
                  Inspection Service",
  howpublished = "\url{https://www.aphis.usda.gov/livestock-poultry-disease/avian/avian-influenza/hpai-detections/livestock/dairy-herd-status-program}",
  note         = "Accessed: 2025-7-11",
  language     = "en"
}

@ARTICLE{Eales2026-op,
  title     = "Modeling of {H5N1} influenza virus kinetics during dairy cattle
               infection suggests the timing of infectiousness",
  author    = "Eales, Oliver and McCaw, James M and Shearer, Freya M",
  journal   = "PLoS Biol.",
  publisher = "Public Library of Science (PLoS)",
  volume    =  24,
  number    =  1,
  pages     = "e3003586",
  month     =  jan,
  year      =  2026,
  language  = "en"
}

@MISC{United-States-Department-of-Agriculture:-National-Agricultural-Statistics-ServiceUnknown-yo,
  title        = "United States 2022 Census of Agriculture. Table 11. Cattle and
                  Calves - Inventory and Sales: 2022 and 2017 (accessed 09
                  December 2024)",
  author       = "{United States Department of Agriculture: National
                  Agricultural Statistics Service}",
  howpublished = "\url{https://www.nass.
                  usda.gov/Publications/AgCensus/2022/Full\_Report/Volume\_1,\_
                  Chapter\_2\_US\_State\_Level/st99\_2\_011\_011.pdf}"
}

@TECHREPORT{noauthor_2026-gj,
  title       = "{USDA} Milk Production Report",
  publisher   = "National Agricultural Statistics Service",
  institution = "United States Department of Agriculture: National Agricultural
                 Statistics Service",
  year        =  2026,
  language    = "en"
}

@MISC{United-Stated-Department-of-Agriculture:-Animal-and-Plant-Health-Inspection-ServiceUnknown-zb,
  title        = "Federal Orders",
  author       = "{United Stated Department of Agriculture: Animal and Plant
                  Health Inspection Service}",
  howpublished = "\url{https://www.aphis.usda.gov/livestock-poultry-disease/avian/avian-influenza/hpai-detections/livestock/federal-order}",
  note         = "Accessed: 2025-7-11",
  language     = "en"
}

@ARTICLE{Sellman2022-as,
  title     = "Modeling {U}.{S}. cattle movements until the cows come home: Who
               ships to whom and how many?",
  author    = "Sellman, Stefan and Beck-Johnson, Lindsay M and Hallman, Clayton
               and Miller, Ryan S and Bonner, Katharine A Owers and Portacci,
               Katie and Webb, Colleen T and Lindström, Tom",
  journal   = "Comput. Electron. Agric.",
  publisher = "Elsevier BV",
  volume    =  203,
  number    =  107483,
  pages     =  107483,
  month     =  dec,
  year      =  2022,
  language  = "en"
}

@ARTICLE{Nguyen2025-bq,
  title     = "Emergence and interstate spread of highly pathogenic avian
               influenza A({H5N1}) in dairy cattle in the United States",
  author    = "Nguyen, Thao-Quyen and Hutter, Carl R and Markin, Alexey and
               Thomas, Megan and Lantz, Kristina and Killian, Mary Lea and
               Janzen, Garrett M and Vijendran, Sriram and Wagle, Sanket and
               Inderski, Blake and Magstadt, Drew R and Li, Ganwu and Diel,
               Diego G and Frye, Elisha Anna and Dimitrov, Kiril M and Swinford,
               Amy K and Thompson, Alexis C and Snekvik, Kevin R and Suarez,
               David L and Lakin, Steven M and Schwabenlander, Stacey and Ahola,
               Sara C and Johnson, Kammy R and Baker, Amy L and Robbe-Austerman,
               Suelee and Torchetti, Mia Kim and Anderson, Tavis K",
  journal   = "Science",
  publisher = "American Association for the Advancement of Science (AAAS)",
  volume    =  388,
  number    =  6745,
  pages     = "eadq0900",
  month     =  apr,
  year      =  2025,
  language  = "en"
}

@ARTICLE{Allen2012-hz,
  title     = "Extinction thresholds in deterministic and stochastic epidemic
               models",
  author    = "Allen, Linda J S and Lahodny, Jr, Glenn E",
  journal   = "J. Biol. Dyn.",
  publisher = "Informa UK Limited",
  volume    =  6,
  number    =  2,
  pages     = "590--611",
  month     =  mar,
  year      =  2012,
  language  = "en"
}

@ARTICLE{Stelwagen2013-kc,
  title     = "Invited review: reduced milking frequency: milk production and
               management implications",
  author    = "Stelwagen, K and Phyn, C V C and Davis, S R and Guinard-Flament,
               J and Pomiès, D and Roche, J R and Kay, J K",
  journal   = "J. Dairy Sci.",
  publisher = "American Dairy Science Association",
  volume    =  96,
  number    =  6,
  pages     = "3401--3413",
  month     =  jun,
  year      =  2013,
  language  = "en"
}

@INCOLLECTION{Reinemann2013-nb,
  title     = "Milking machines and milking parlors",
  author    = "Reinemann, Douglas J",
  booktitle = "Handbook of Farm, Dairy and Food Machinery Engineering",
  publisher = "Elsevier",
  pages     = "177--197",
  month     =  jan,
  year      =  2013
}

@TECHREPORT{Dairy-Australia2020-nv,
  title       = "Farm guidelines for mastitis control",
  author      = "{Dairy Australia}",
  institution = "Dairy Australia",
  year        =  2020,
  language    = "en"
}

@ARTICLE{Facciuolo2025-wy,
  title     = "Dairy cows develop protective immunity against reinfection with
               bovine {H5N1} influenza virus",
  author    = "Facciuolo, Antonio and Aubrey, Lauren and Barron-Castillo, Ulises
               and Berube, Nathalie and Norleen, Carla and McCreary, Shannon and
               Huang, Yanyun and Pessoa, Natalia and Jacome, Leslie Macas and
               Mubareka, Samira and McGeer, Allison and Berhane, Yohannes and
               Gerdts, Volker and Van Kessel, Andrew and Warner, Bryce and Zhou,
               Yan",
  journal   = "Nat. Microbiol.",
  publisher = "Springer Science and Business Media LLC",
  volume    =  10,
  number    =  6,
  pages     = "1366--1377",
  month     =  jun,
  year      =  2025,
  language  = "en"
}

@ARTICLE{Pena-Mosca2025-lk,
  title     = "The impact of highly pathogenic avian influenza {H5N1} virus
               infection on dairy cows",
  author    = "Peña-Mosca, Felipe and Frye, Elisha A and MacLachlan, Matthew J
               and Rebelo, Ana Rita and de Oliveira, Pablo S B and Nooruzzaman,
               Mohammed and Koscielny, Melanie Prarat and Zurakowski, Michael
               and Lieberman, Zoe R and Leone, William M and Elvinger, François
               and Nydam, Daryl V and Diel, Diego G",
  journal   = "Nat. Commun.",
  publisher = "Springer Science and Business Media LLC",
  volume    =  16,
  number    =  1,
  pages     =  6520,
  month     =  jul,
  year      =  2025,
  language  = "en"
}

@ARTICLE{Kaiser2025-zh,
  title    = "Highly pathogenic avian influenza A({H5N1}) virus stability in
              irradiated raw milk and wastewater and on surfaces, United States",
  author   = "Kaiser, Franziska and Cardenas, Santiago and Yinda, Kwe Claude and
              Mukesh, Reshma K and Ochwoto, Missiani and Gallogly, Shane and
              Wickenhagen, Arthur and Bibby, Kyle and de Wit, Emmie and Morris,
              Dylan and Lloyd-Smith, James O and Munster, Vincent J",
  journal  = "Emerg. Infect. Dis.",
  volume   =  31,
  number   =  4,
  pages    = "833--837",
  month    =  apr,
  year     =  2025,
  language = "en"
}

@ARTICLE{Lee2025-hs,
  title    = "Dairy cows infected with influenza A({H5N1}) reveals low
              infectious dose and transmission barriers",
  author   = "Lee, Carolyn and Tarbuck, Natalie N and Cochran, Hannah J and
              Foreman, Bryant M and Boley, Patricia and Khatiwada, Saroj and
              Dhakal, Alok and Adefaye, Khadijat O and Schrock, Jennifer and
              Jahid, Mohammad Jawad and Laocharoensuk, Thamonpan and Suresh,
              Raksha and Shekoni, Olaitan and Stevens, Erika and Dolatyabi, Sara
              and Sanders, Christina and Ohl, Elizabeth and Huey, Devra and
              Hanson, Juliette and Gourapura, Renukaradhya and Webby, Richard J
              and Warren, Cody J and Kenney, Scott P and Bowman, Andrew S",
  journal  = "Research Square",
  month    =  jun,
  year     =  2025,
  language = "en"
}

@ARTICLE{Cabezas2021-ij,
  title     = "Spatial and network analysis of {U}.{S}. livestock movements
               based on Interstate Certificates of Veterinary Inspection",
  author    = "Cabezas, A H and Sanderson, M W and Lockhart, C Y and Riley, K A
               and Hanthorn, C J",
  journal   = "Prev. Vet. Med.",
  publisher = "Elsevier BV",
  volume    =  193,
  number    =  105391,
  pages     =  105391,
  month     =  aug,
  year      =  2021,
  language  = "en"
}

@ARTICLE{Crespo-Bellido2026-at,
  title       = "Emergence of {D1}.1 reassortant {H5N1} avian influenza viruses
                 in North America",
  author      = "Crespo-Bellido, Alvin and Trovão, Nídia S and Puryear, Wendy
                 and Maksiaev, Alexander and Pekar, Jonathan E and Baele, Guy
                 and Dellicour, Simon and Nelson, Martha I",
  journal     = "bioRxivorg",
  institution = "bioRxiv",
  pages       = "2025.12.19.695329",
  month       =  apr,
  year        =  2026,
  language    = "en"
}

@ARTICLE{Campbell2026-af,
  title     = "Surveillance on California dairy farms reveals multiple possible
               sources of {H5N1} influenza virus transmission",
  author    = "Campbell, A J and Shephard, Meredith and Paulos, Abigail P and
               Pauly, Matthew D and Vu, Michelle N and Stenkamp-Strahm, Chloe
               and Bushfield, Kaitlyn and Hunter-Binns, Betsy and Sablon,
               Orlando and Bendall, Emily E and Fitzimmons, William J and
               Brizuela, Kayla and Quirk, Grace E and Kumar, Nirmal and
               McCluskey, Brian and Shetty, Nishit and Marr, Linsey C and
               Guthmiller, Jenna J and Santos, Jefferson J S and Hensley, Scott
               E and Marshall, Edith S and Abernathy, Kevin and Lauring, Adam S
               and Melody, Blaine T and Wolfe, Marlene K and Lombard, Jason and
               Lakdawala, Seema S",
  journal   = "PLoS Biol.",
  publisher = "Public Library of Science (PLoS)",
  volume    =  24,
  number    =  5,
  pages     = "e3003761",
  month     =  may,
  year      =  2026,
  language  = "en"
}
\pagebreak
\section*{Supplementary figures}

\setcounter{figure}{0}
\setcounter{table}{0}
\let\oldthefigure\thefigure
\renewcommand{\thefigure}{S\oldthefigure}
\renewcommand{\thetable}{S\arabic{table}}

\pagebreak

\begin{table}
    \centering
    \includegraphics[height=0.7\linewidth, angle=90]{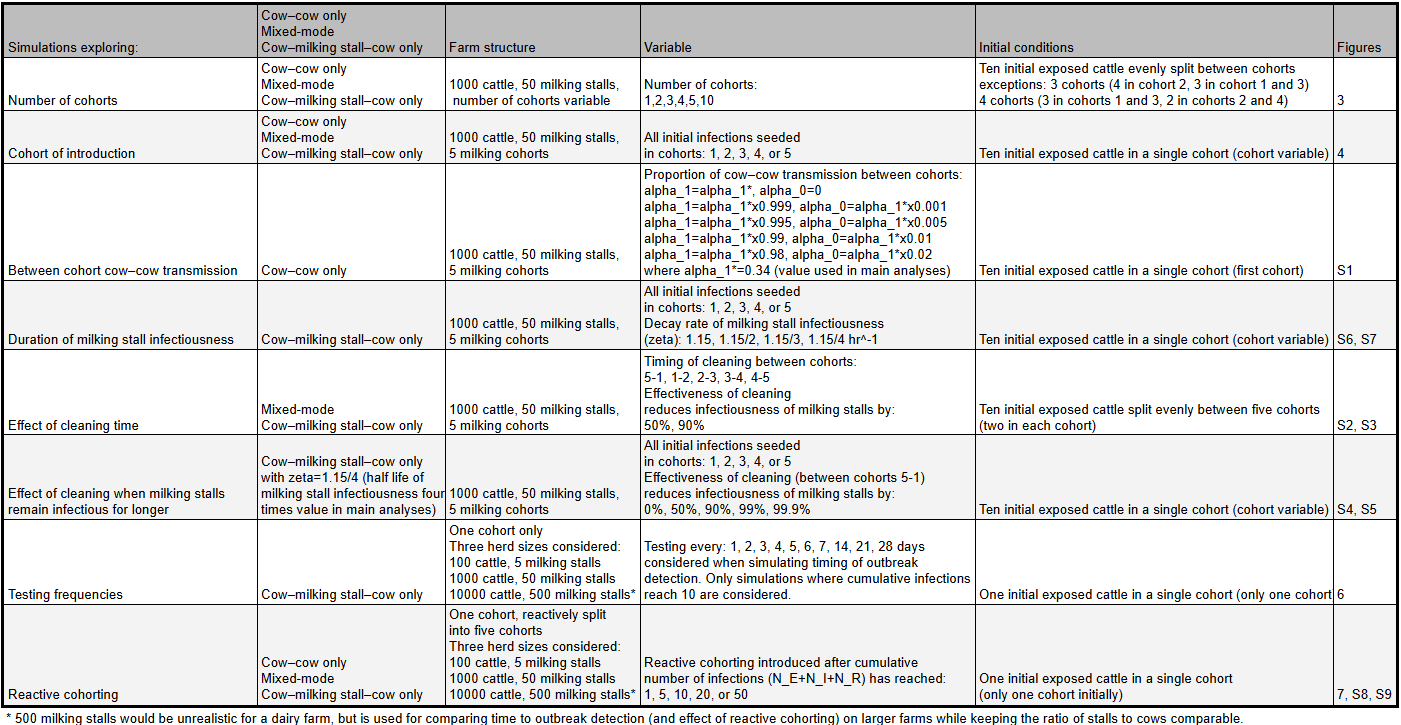}
    \caption{Details of all simulation studies performed}
    \label{tab:sims_table}
\end{table}

\begin{figure}
    \centering
    \includegraphics[width=0.5\linewidth]{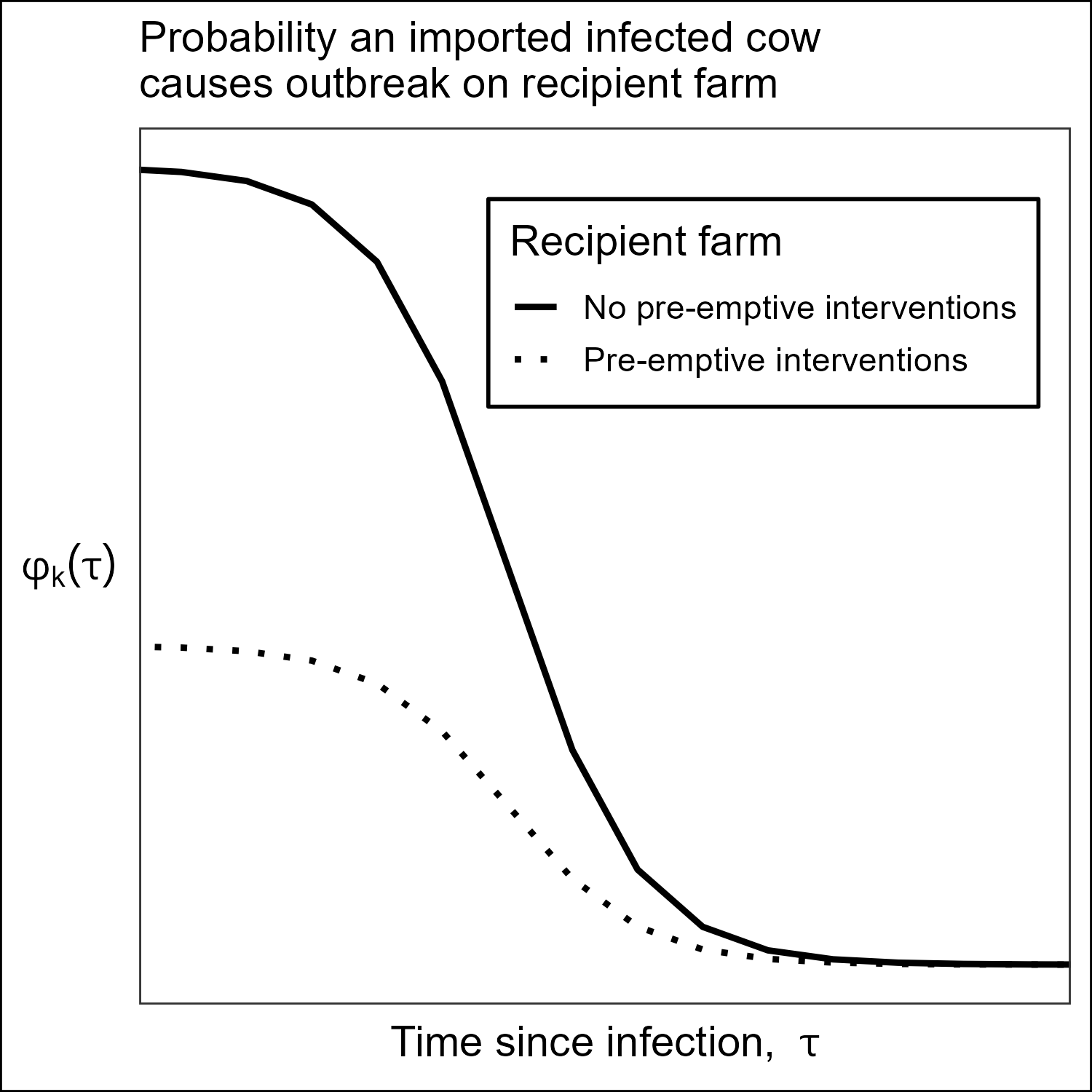}
    \caption{Graphical representation of the potential impact of pre-emptive interventions on the probability an introduced infected cow causes an outbreak. Pre-emptive measures on the recipient farm (dotted lines), may reduce the probability an outbreak occurs following the importation of an infected animal. This can reduce the overall probability of an outbreak on one farm resulting in an outbreak on another farm (see Figure \ref{fig:propagation-of-outbreaks}).}
    \label{fig:propagation-of-outbreaks-phi-sup}
\end{figure}

\begin{figure}
    \centering
    \includegraphics[width=1.0\linewidth]{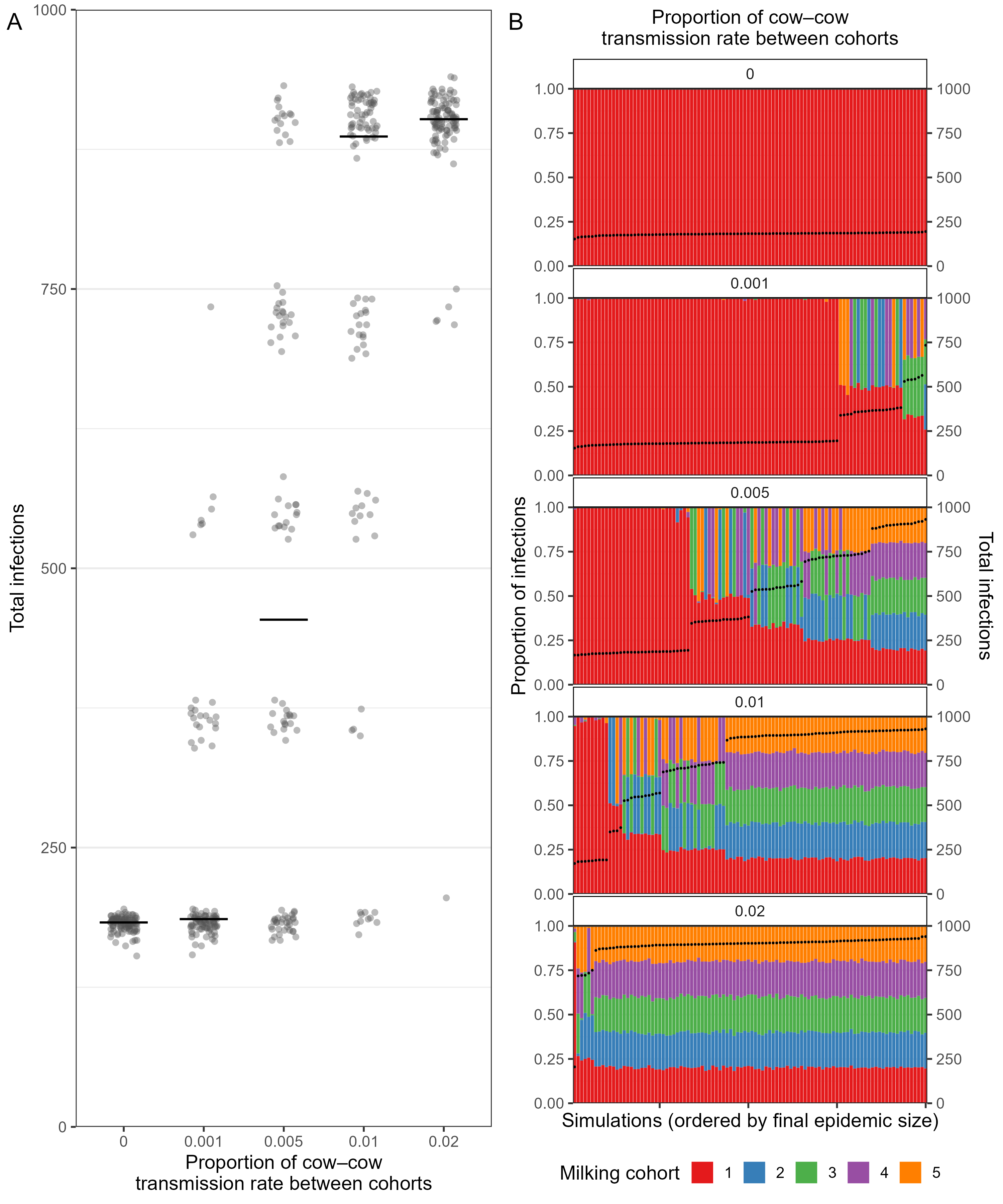}
    \caption{ Sensitivity of cow--cow transmission regime to between cohort transmission. (A) The total outbreak size (i.e. the total number of infections) across 100 simulations seeded with ten initial infections ($N_E(t=0)=10$) in cohort 1, for the cow--cow transmission only regime with different levels of between cohort transmission (relative to within cohort transmission). For all scenarios the median outbreak size across 100 simulations is also shown (black line) (B) The overall distribution of infections across all simulations in A. Stacked bars show the proportion of infections in each cohort (colours, left-hand y-axis) across all simulations. The simulations are ordered from smallest outbreak to largest outbreak, with the total outbreak size also plotted (points, right-hand y-axis).
    }
    \label{fig:sens_bct}
\end{figure}

\begin{figure}
    \centering
    \includegraphics[width=1.0\linewidth]{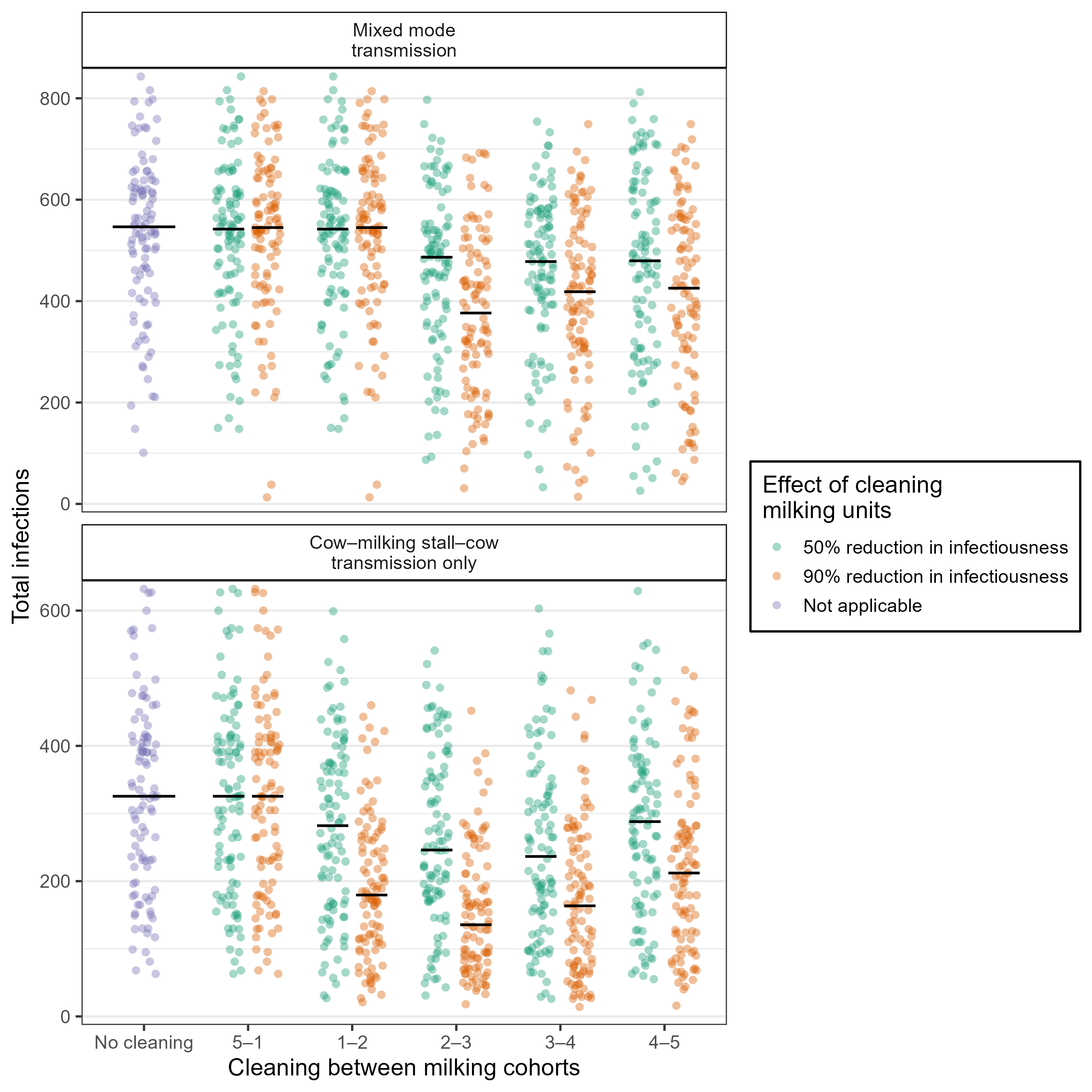}
    \caption{The effect of enhanced cleaning of milking stalls for parameters used in main analyses. The total outbreak size (i.e. the total number of infections) across 100 simulations seeded with ten initial infections ($N_E(t=0)=10$) seeded evenly across five cohorts (two in each cohort), for two transmission regimes (panels), different effectiveness of cleaning milking stalls (colour) and with cleaning done once between a different pair of milking cohorts (x-axis). For all scenarios the median outbreak size across 100 simulations is also shown (black line). Note that cleaning between milking cohorts 5 and 1 represents cleaning between milking periods. 
    }
    \label{fig:sens_cleaning1}
\end{figure}

\begin{figure}
    \centering
    \includegraphics[width=1.0\linewidth]{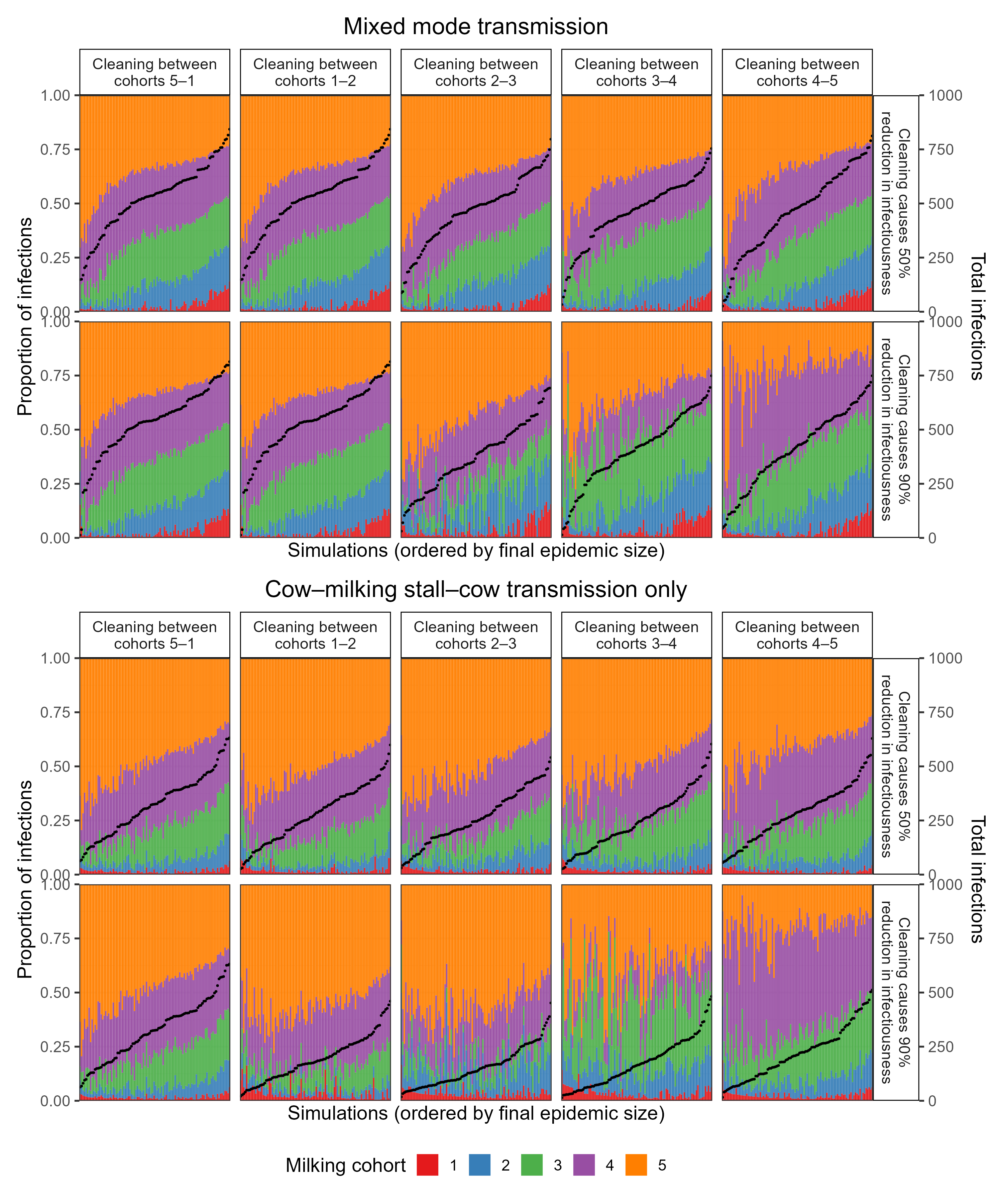}
    \caption{The effect of enhanced cleaning of milking stalls by milking cohorts for parameters used in the main analyses.    
    The overall distribution of infections across all simulations in Figure \ref{fig:sens_cleaning1}. Stacked bars show the proportion of infections in each cohort (colours, left-hand y-axis) across all simulations. The simulations are ordered from smallest outbreak to largest outbreak, with the total outbreak size also plotted (points, right-hand y-axis). Panels are divided by transmission regimes, which pair of cohorts cleaning occurs between, and what is the effectiveness of cleaning milking stalls.  }
    \label{fig:sens_cleaning2}
\end{figure}

\begin{figure}
    \centering
    \includegraphics[width=1.0\linewidth]{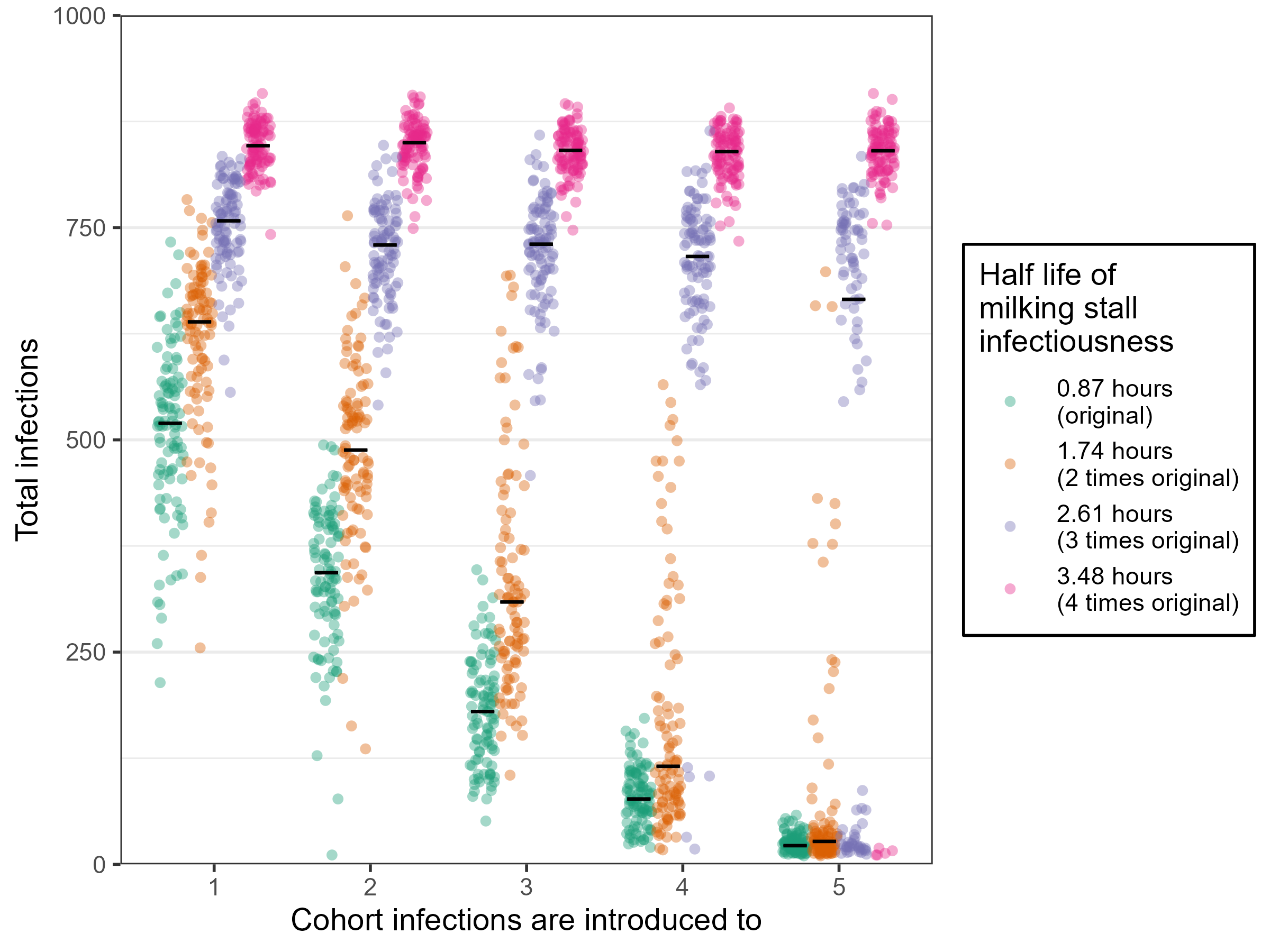}
    \caption{ 
    The effect of cohort in which infections are introduced when milking stalls remain infectious for longer for the cow--milking stall--cow transmission regime. The total outbreak size (i.e. the total number of infections) across 100 simulations seeded with ten initial infections ($N_E(t=0)=10$), for the cow--cow transmission regime with different assumed durations of infectiousness (i.e. different values of $\zeta$, the decay rate of infectiousness) for milking stalls (colours) and with all infections seeded into a different milking cohort (five milking cohorts in simulations). For all scenarios the median outbreak size across 100 simulations is also shown (black line).
    }
    \label{fig:sens_dt1}
\end{figure}

\begin{figure}
    \centering
    \includegraphics[width=1.0\linewidth]{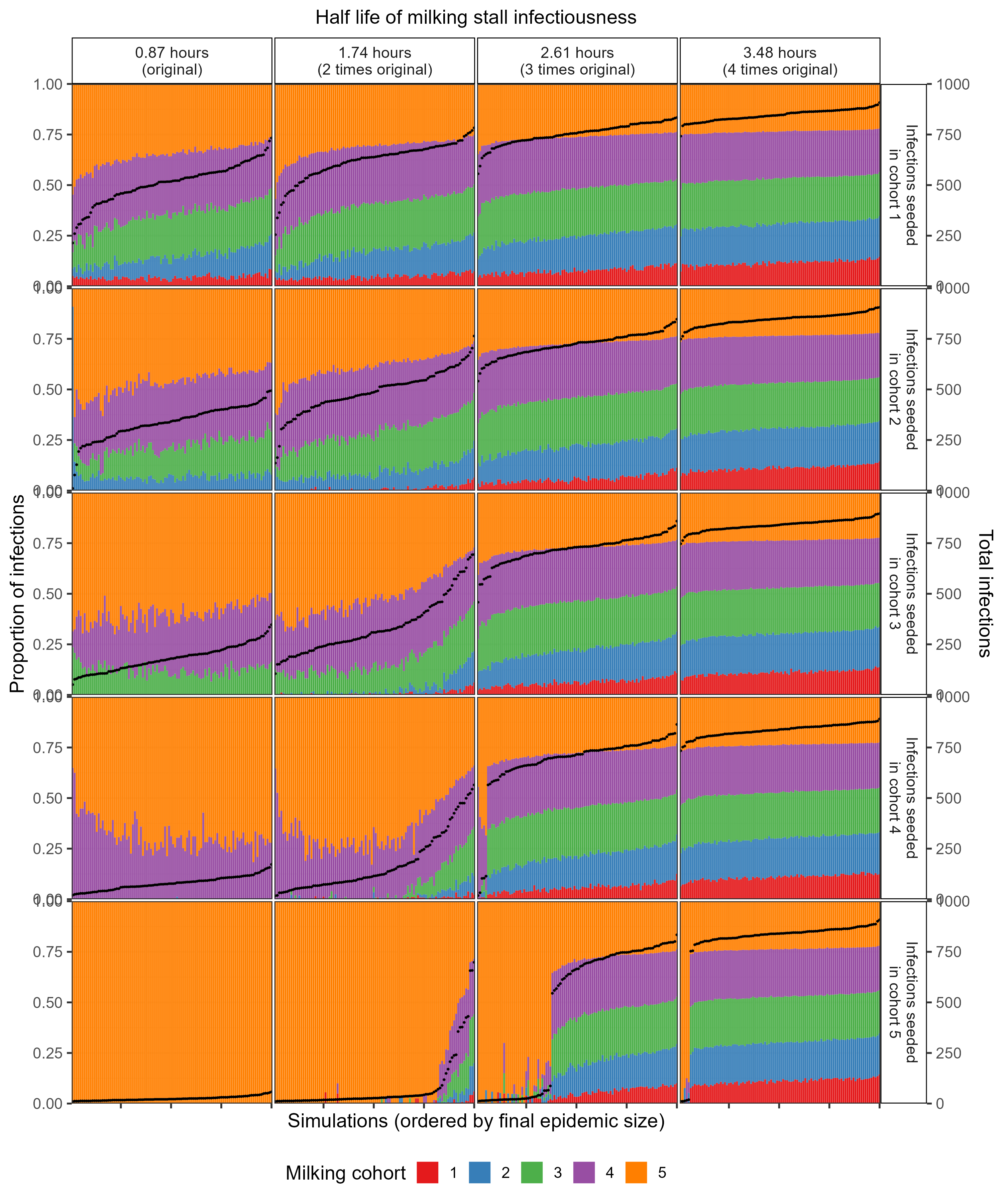}
    \caption{ The effect of cohort in which infections are introduced by milking stall when milking stalls remain infectious for longer for the cow--milking stall--cow transmission regime. The overall distribution of infections across all simulations in Figure \ref{fig:sens_dt1}. Stacked bars show the proportion of infections in each cohort (colours, left-hand y-axis) across all simulations. The simulations are ordered from smallest outbreak to largest outbreak, with the total outbreak size also plotted (points, right-hand y-axis). Panels are divided by the half life of milking stall infectiousness (i.e. different values of, $\zeta$, the decay rate of infectiousness), and the cohort in which infections are introduced.
    }
    \label{fig:sens_dt2}
\end{figure}

\begin{figure}
    \centering
    \includegraphics[width=1.0\linewidth]{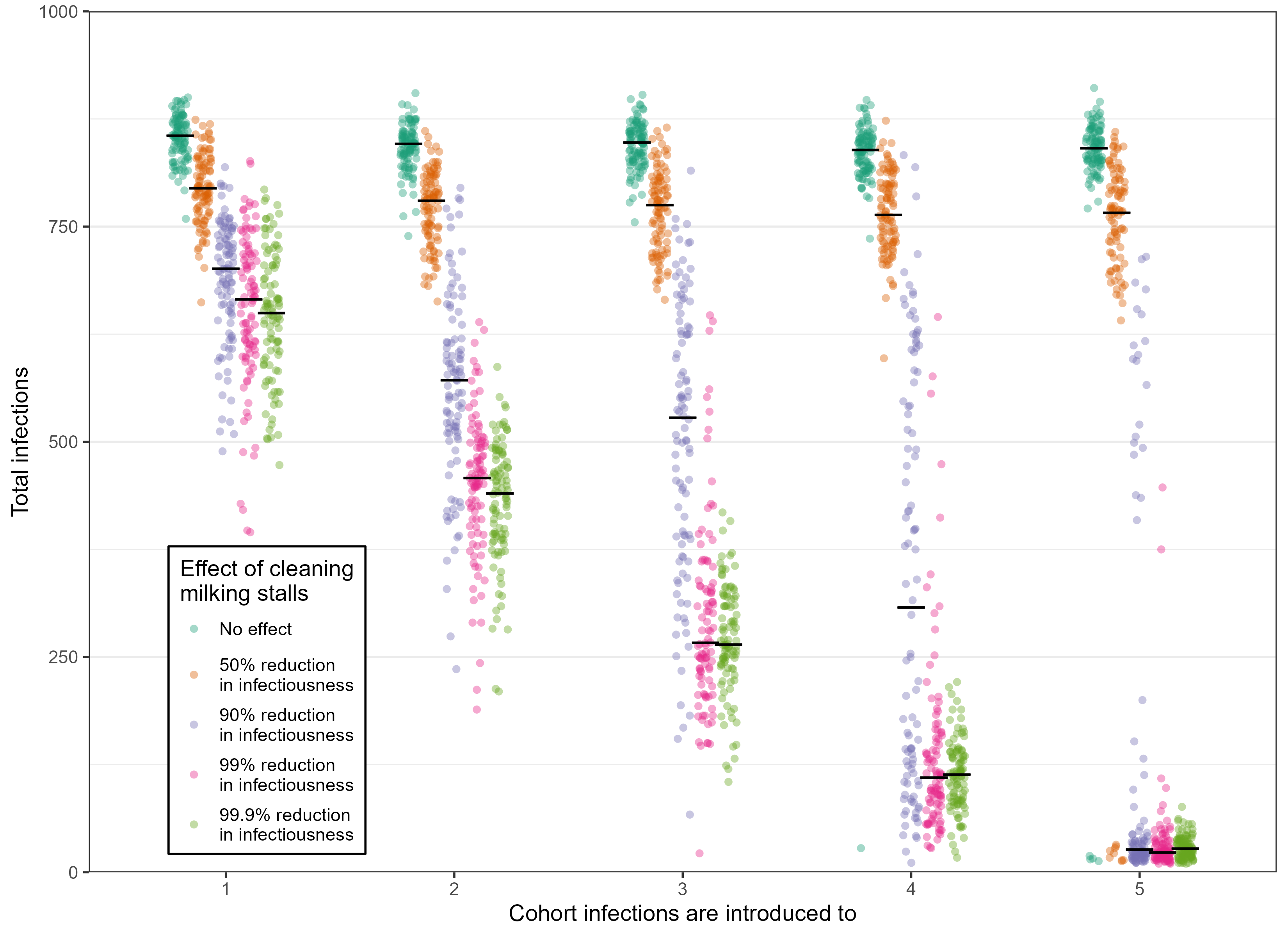}
    \caption{ The effect of cohort in which infections are introduced when milking stalls remain infectious for longer but cleaning is performed between milking periods. The total outbreak size (i.e. the total number of infections) across 100 simulations seeded with ten initial infections ($N_E(t=0)=10$), for the cow--milking stall--cow transmission regime with the half life of milking stall infectiousness assumed to be four times the original (i.e. $\zeta$, the decay rate of infectiousness, is a quarter the value used in main analyses), for different effectiveness of cleaning milking stalls (colours) and with all infections seeded into a different milking cohort (five milking cohorts in simulations). For all scenarios the median outbreak size across 100 simulations is also shown (black line).
    }
    \label{fig:sens_dt2_cleaning1}
\end{figure}

\begin{figure}
    \centering
    \includegraphics[width=1.0\linewidth]{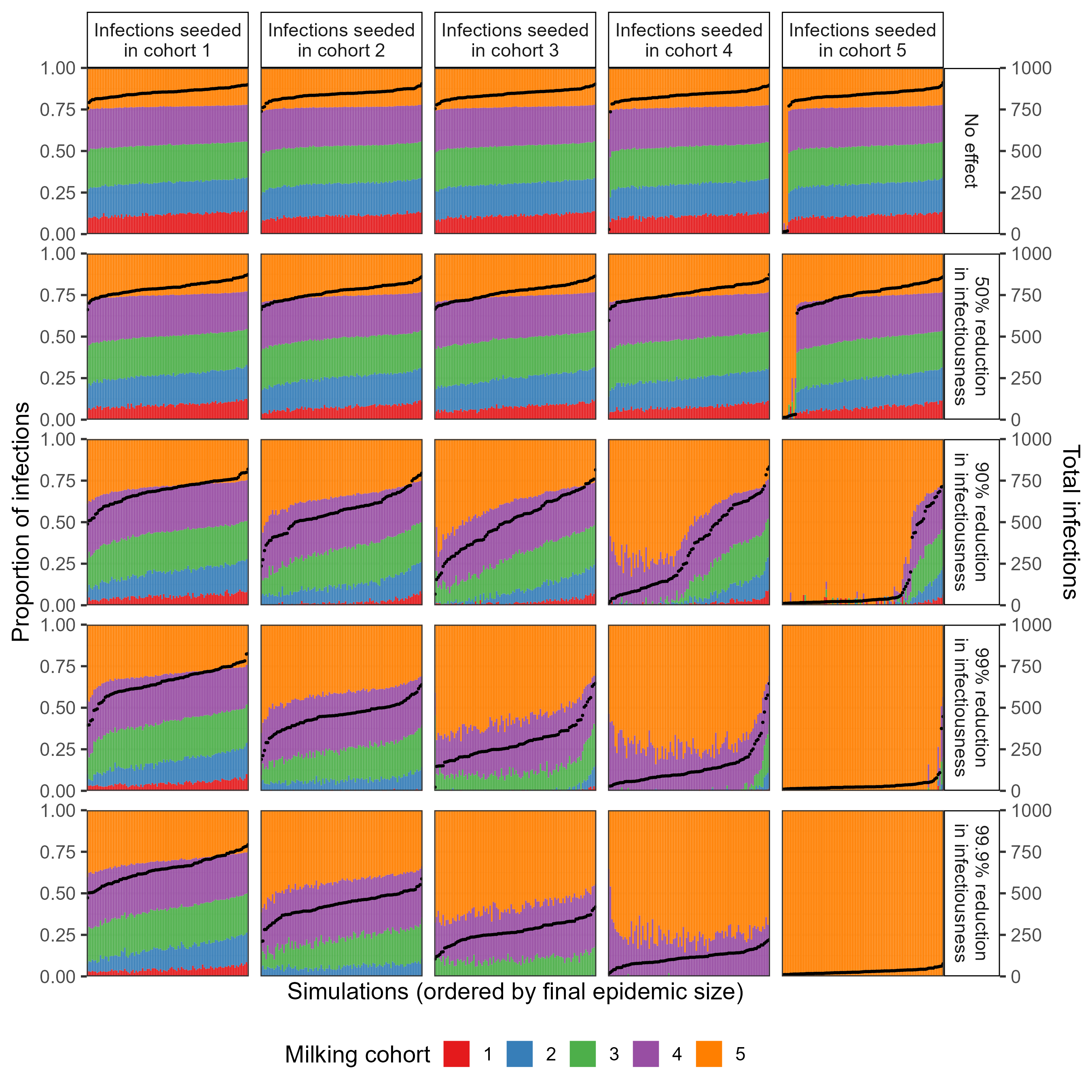}
    \caption{ 
     The effect of cohort in which infections are introduced by milking stall when milking stalls remain infectious for longer but cleaning is performed between milking periods. The overall distribution of infections across all simulations in Figure \ref{fig:sens_dt2_cleaning1}. Stacked bars show the proportion of infections in each cohort (colours, left-hand y-axis) across all simulations. The simulations are ordered from smallest outbreak to largest outbreak, with the total outbreak size also plotted (points, right-hand y-axis). Panels are divided by the effectiveness of cleaning milking stalls, and the cohort in which infections are introduced.
     }
    \label{fig:sens_dt2_cleaning2}
\end{figure}

\begin{figure}
    \centering
    \includegraphics[width=1.0\linewidth]{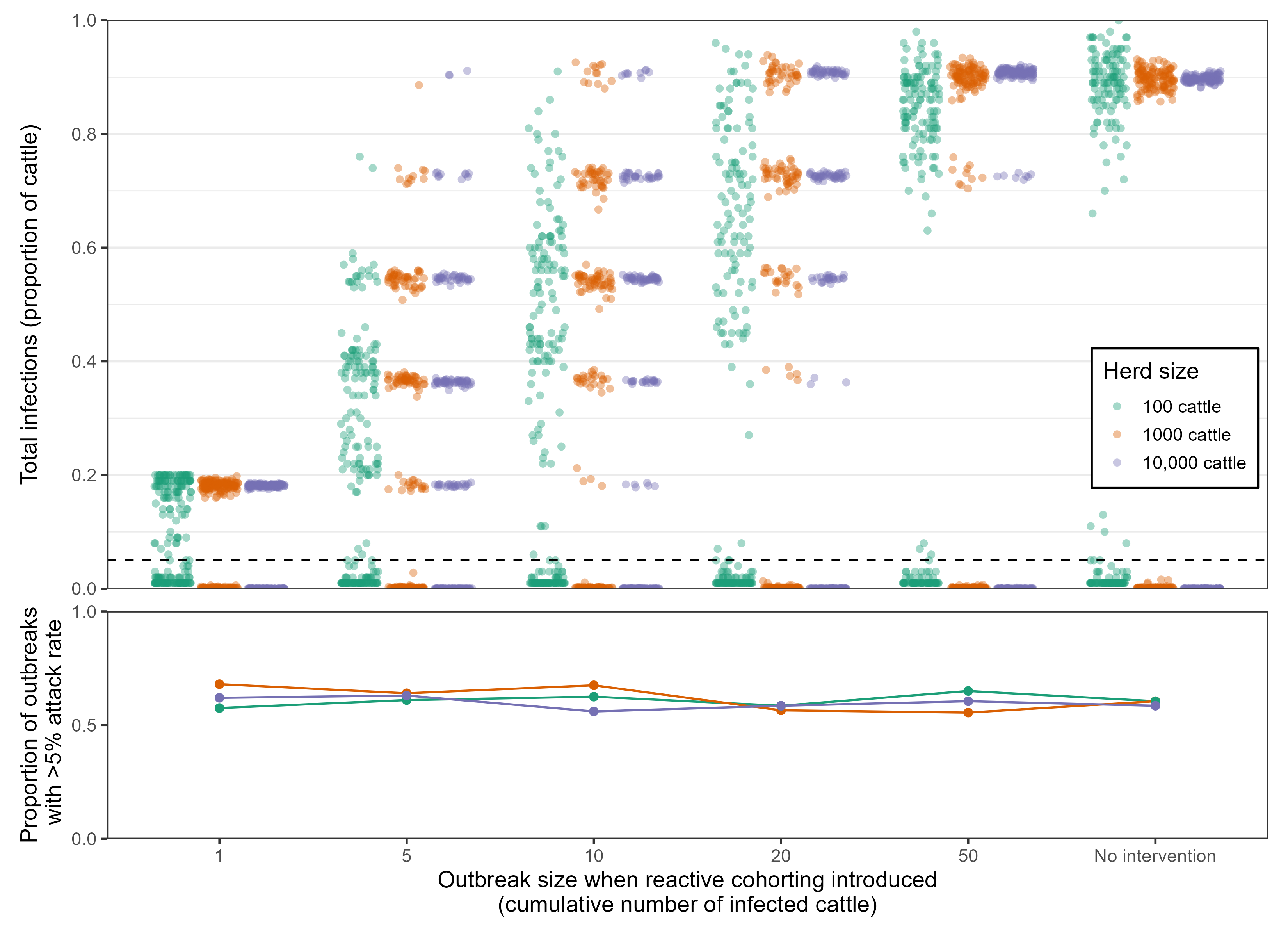}
    \caption{
    The effect of reactively cohorting at different thresholds during an outbreak for the cow--cow transmission regime. (Top) The total outbreak size (i.e. the total proportion of infected cattle) across 200 simulations seeded with one initial infection ($N_E(t=0)=1$) for the cow--cow transmission regime for: different herd sizes (colours); and different thresholds of cumulative infections at which reactive cohorting (five cohorts) is performed. Note there are many simulations where outbreaks are small even when no intervention is performed; this is because the outbreak was only seeded with one initial infection and so many simulations will go extinct before an outbreak occurs, and the threshold for reactive cohorting might not be met. (Bottom). The proportion of simulations for which the outbreak size was less than 5\% (black dashed line in top panel) for the different thresholds at which reactive cohorting is performed (points) and for the different herd sizes (colours).}
    \label{fig:reactive_cohorts_param1}
\end{figure}

\begin{figure}
    \centering
    \includegraphics[width=1.0\linewidth]{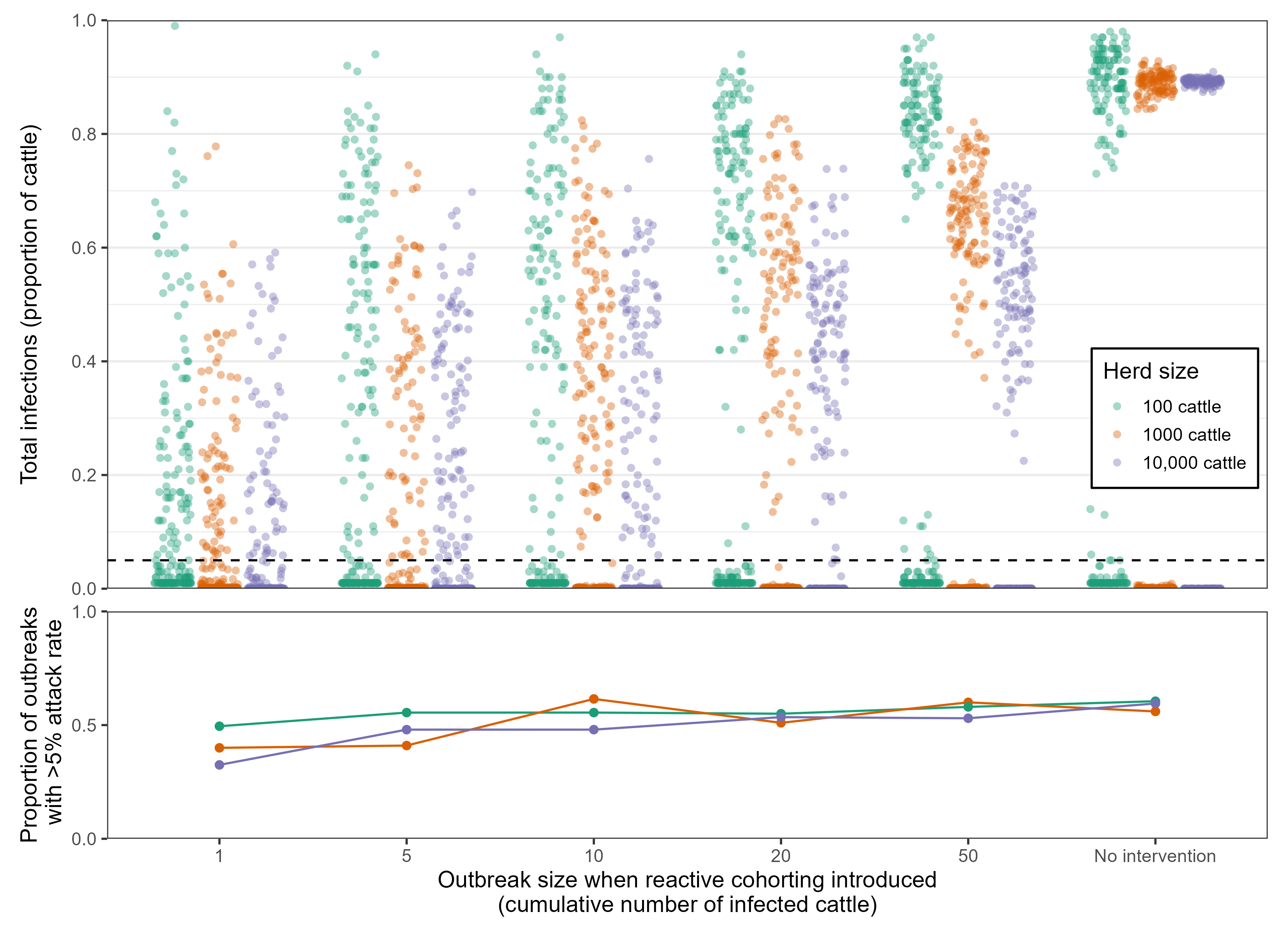}
    \caption{  The effect of reactively cohorting at different thresholds during an outbreak for the mixed-mode transmission regime. (Top) The total outbreak size (i.e. the total proportion of infected cattle) across 200 simulations seeded with one initial infection ($N_E(t=0)=1$) for the mixed-mode transmission regime for: different herd sizes (colours); and different thresholds of cumulative infections at which reactive cohorting (five cohorts) is performed. Note there are many simulations where outbreaks are small even when no intervention is performed; this is because the outbreak was only seeded with one initial infection and so many simulations will go extinct before an outbreak occurs, and the threshold for reactive cohorting might not be met. (Bottom). The proportion of simulations for which the outbreak size was less than 5\% (black dashed line in top panel) for the different thresholds at which reactive cohorting is performed (points) and for the different herd sizes (colours).
    }
    \label{fig:reactive_cohorts_param3}
\end{figure}

\end{document}